\documentclass{pasa}%

\usepackage{graphicx}
\usepackage[dvipsnames]{xcolor}

\title[RFI in the FM band at the MRO]{A survey of spatially and temporally resolved radio frequency interference in the FM band at the Murchison Radio-astronomy Observatory}

\author[Tingay et al.]{Tingay, S.J., Sokolowski, M., Wayth, R., \& Ung, D.
\affil{International Centre for Radio Astronomy Research, Curtin University, Bentley, WA 6102, Australia}%
}%

\jid{PASA}
\doi{10.1017/pas.\the\year.xxx}
\jyear{\the\year}

\usepackage{aas_macros}
\usepackage{hyperref} 
\hypersetup{colorlinks,citecolor=blue,linkcolor=blue,urlcolor=blue}

\begin{document}

\begin{frontmatter}
\maketitle

\begin{abstract}
We present the first survey of radio frequency interference (RFI) at the future site of the low frequency Square Kilometre Array (SKA), the Murchison Radio-astronomy Observatory (MRO), that both temporally and spatially resolves the RFI.  The survey is conducted in a 1 MHz frequency range within the FM band, designed to encompass the closest and strongest FM transmitters to the MRO (located in Geraldton, approximately 300 km distant).  Conducted over approximately three days using the second iteration of the Engineering Development Array in an all-sky imaging mode, we find a range of RFI signals.  We are able to categorise the signals into: those received directly from the transmitters, from their horizon locations; reflections from aircraft (occupying approximately 13\% of the observation duration); reflections from objects in Earth orbit; and reflections from meteor ionisation trails.  In total we analyse 33,994 images at 7.92 s time resolution in both polarisations with angular resolution of approximately 3.5$^{\circ}$, detecting approximately forty thousand RFI events.  This detailed breakdown of RFI in the MRO environment will enable future detailed analyses of the likely impacts of RFI on key science at low radio frequencies with the SKA.
\end{abstract}

\end{frontmatter}

\section{INTRODUCTION }
\label{sec:int}

The next generation of radio telescopes, built to explore the evolution of the Universe to very high redshift, during the so-called Epoch of Reionisation (EoR; \citet{2006PhR...433..181F}), will be located at sites selected to have very low levels of human-made Radio Frequency Interference (RFI).

For example, the low frequency component of the Square Kilometre Array (SKA\footnote{www.skatelescope.org}) will be built at the Murchison Radio-astronomy Observatory (MRO), located in the Mid West of Western Australia.  The MRO is also home to two SKA Precursor telescopes, CSIRO's ASKAP \citep{2008ExA....22..151J} and the Murchison Widefield Array (MWA) \citep{2018PASA...35...33W,2013PASA...30....7T}.  The MWA remains the only fully operational Precursor for the low frequency SKA and has been an operational facility since 2013.

With a frequency range of approximately 70 - 300 MHz, the MWA spans a number of Earth and space-based broadcast bands, including the ubiquitous FM band (approximately 88 - 108 MHz in Australia), constituting a primary source of RFI at the MRO.  Likewise, the frequency range for SKA\_low is 50 - 350 MHz, also encompassing the FM band.

The effect of RFI on key science programs at low frequencies, including EoR experiments, is very significant.  Aside from locating telescopes such as the MWA at the best sites, such as the MRO, substantial general effort has been put into the identification and removal of RFI from radio telescope data (RFI mitigation).  An excellent assessment of the effects of RFI on EoR measurements, plus a review of the general considerations for RFI mitigation, can be found in \citet{2020arXiv200407819W}; they place limits on the tolerable RFI budget before EoR measurements are adversely affected.  
However, the practical realistion of an allowable RFI budget is complex.  \citet{2020arXiv200407819W} find that a detailed knowledge of the spatial and temporal distribution of the RFI, the duration of observations, the characteristics of the telescope in question, and the detailed scheduling of the telescope (where it is pointing and when) are required.  Measurements that attempt to characterise the RFI environment in which a telescope operates are therefore very important in order to understand the overall performance of key science observations, such as for the EoR experiment.

Previously, a study of the RFI environment of the MRO, utilising the MWA over 10 nights of observations in the frequency range 72 - 231 MHz, was conducted by \citet{2015PASA...32....8O}.  Using their methods of RFI detection and excision, they found that 1.1\% of data required excision, but that all frequency ranges remained highly usable, including in the FM and digital TV bands.

More recently, \citet{2016arXiv161004696S} and \citet{2017rfi..confE...1S} examined RFI statistics for the MRO in the frequency range 70 - 300 MHz using the BIGHORNS global EoR signal experiment \citep{2015PASA...32....4S}.  They find examples of ducting events caused by atmospheric conditions that propagate signals from distant transmitters to the MRO.  A similar, highly comprehensive survey of the LOFAR RFI environment was conducted by \citet{2013A&A...549A..11O}.

These previous studies of the RFI environment of the MRO at low frequencies have collected information as a function of frequency and of time, but not as a function of location on the celestial sphere.  The previous studies have obtained information that has been integrated spatially prior to analysis.  In this paper, we examine the RFI characteristics of the MRO for one frequency within the FM band, resolved both temporally and spatially.  A range of RFI signals are expected at low radio frequencies, with the FM band one of the most prominent in the SKA\_low frequency range.  At these frequencies, as well as ducted propagation from distant transmitters, multi-path reception of RFI is expected due to reflections off objects within the environment.  For example, reflections off aircraft \citep{2020arXiv200407819W}, ionised meteor trails \citep{2018MNRAS.477.5167Z}, and satellites \citep{2020PASA...37...10P, 2018MNRAS.477.5167Z, Tingay2013OnFeasibility} are expected.  Such signals are ubiquitous and have been studied in detail for other low frequency radio telescopes such as the Owens Valley Long Wavelength Array \citep{https://doi.org/10.7907/25dp-j474}.  These reflected signals are expected to be localised in time and direction and contribute at different levels to the total RFI budget.  We aim to assess and characterise these different contributions here, presenting the information required for a range of future quantitative studies of the impact of RFI on key low frequency science for the SKA.

We obtain temporally and spatially resolved RFI measurements by using the second realisation of the so-called Engineering Development Array (EDA2: Wayth et al. 2020, in preparation), an SKA\_low station configuration composed of 256 individual MWA antennas on a 35 m station footprint.

In \S \ref{subsec:obs}, we describe the EDA2 instrument, its characteristics, and the observations undertaken for this work.  In \S \ref{subsec:imcal}, we describe the imaging and calibration of the EDA2 data undertaken to provide the raw images used for further analysis.  In \S \ref{subsec:ann} we describe the data processing undertaken to separate the different categories of RFI signals, and in \S \ref{sec:res} we describe the results, assessing the components of the RFI budget in this frequency band.  Finally, in \S \ref{sec:dis} we discuss the likely impacts of these components of the RFI budget for some areas of key low frequency science.

\section{OBSERVATIONS AND DATA PROCESSING}
\label{sec:obs}

\subsection{Observations}
\label{subsec:obs}

\begin{figure}[h!]
\begin{center}
\includegraphics[width=0.33\textwidth,angle=270]{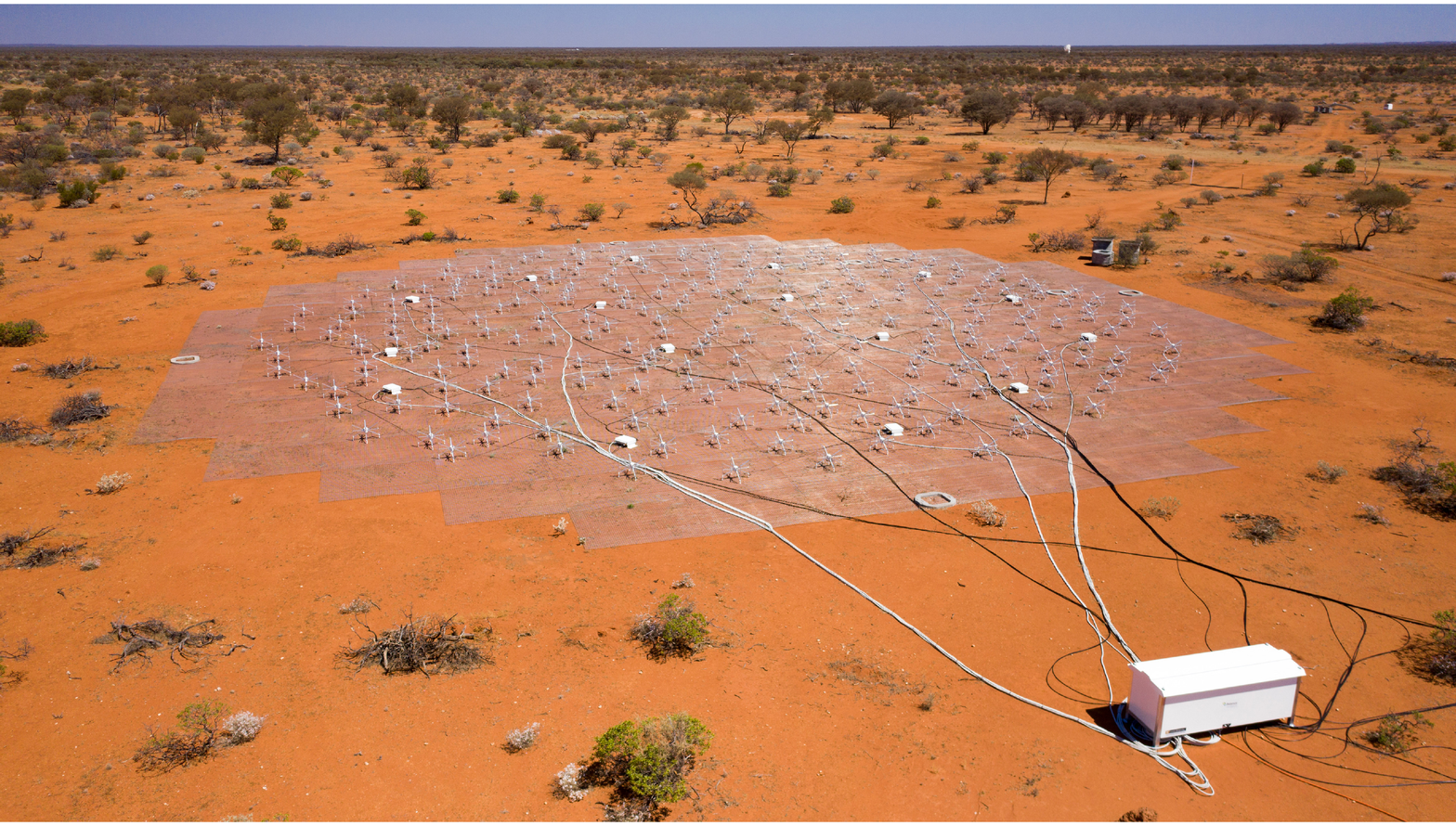}
\caption{An aerial view of the EDA2 instrument used for this work.}
\label{eda}
\end{center}
\end{figure}

Observations were conducted using the Engineering Development Array, version 2 (EDA2: Figure \ref{eda}), an array of 256 MWA antennas arranged in an SKA\_low station configuration (station diameter of 35 m) and located at the MRO (Wayth et al. in preparation).  Analogue signal chains from the individual antennas (both polarisations; X and Y) are digitised and coarse-channalised in the firmware \citep{2017JAI.....641015C} implemented in Tile Processing Modules \citep[TPM;][]{2017JAI.....641014N}. The coarse-channelised voltage streams are received on the data acquisition computer and signals from individual antennas are correlated using the \textsc{xGPU} \citep{xGPU} software correlator.  Thus, the EDA2 forms a 256 element, dual-polarisation interferometer that can be used to form all-sky images using standard interferometric calibration and imaging techniques.

EDA2 data were collected between 2020-01-31 06:30 and 2020-02-03 07:20 UTC at 1.98\,s temporal resolution, as part of commissioning activities for the array.  During these observations, 54 antennas were non-functional for commissioning, leaving a usable array of 202 antennas.  Data were collected over a 0.926\,MHz bandwidth (single coarse channel), at a central frequency of 98.4375\,MHz. This particular coarse channel covered two FM stations from Geraldton at 98.1 and 98.9\,MHz providing excellent data to perform the analysis presented in this paper. The correlation products were saved in \textsc{HDF5} format\footnote{https://www.hdfgroup.org/}, which is envisaged as the data format for the SKA telescope.

\subsection{Imaging and calibration}
\label{subsec:imcal}

The data were converted from native \textsc{HDF5} format into UV FITS files \citep{uvfits}, averaging four consecutive raw integrations into 7.92\,s files, using the zenith as the phase centre for each file. This generated 33,994 individual files to be imaged.
Further processing was performed with the \textsc{Miriad} data processing suite \citep{1995ASPC...77..433S}.

Calibration (phase and amplitude) was performed using the \textsc{mfcal} task on a subset of data from the solar transit and a flux density model of the quiet sun \citep{sun_model}. Baselines shorter than $5\lambda$ were excluded to minimise the contribution from Galactic extended emission.
The calibration was transferred to all datasets, then each file was imaged with \textsc{invert} into $128\times128$ pixel images using robust=-0.5 weights, and deconvolved with \textsc{clean} using 10000 iterations, speed=-1, and phat=0.1. Each image covers the entire hemisphere visible for that file with 1.25 degree pixel scale at the zenith in sine projection, with an angular resolution of $\sim3.5^{\circ}$ at zenith.
The two independent polarisations (corresponding to the `XX' or east-west oriented dipoles and the `YY' or north-south oriented dipoles) were imaged independently.
This process generated two zenith-centred images for each visibility file, which were exported as FITS images. 

\subsection{Radio frequency interference analysis}
\label{subsec:ann}

The FITS images at the 33,994 individual time steps (at each XX and YY polarisation, for a total of 67,988 images) were processed using a Python script.

Each image was loaded from disk and a difference image was formed via subtraction of the previous image in time.  Difference images remove the component of the sky brightness distribution that remains constant from time step to time step, including astronomical sources such as the very bright Galactic Plane.  Difference images reveal signals that change with time between images and are sensitive to time-varying sources of RFI.  Since there are no sources of FM RFI in the immediate vicinity of the MRO, all sources of RFI in this band are due to complex propagation or scattering effects, and are therefore strongly time variable.

The difference images were searched for point sources, using the DAOStarFinder task within the photutils Python module \citep{Bradley_2019_2533376}, based on DAOPHOT \citep{1987PASP...99..191S}.  At the angular resolution of EDA2, any time-variable sources of RFI will be point-like in nature, even scattering objects in motion (aircraft and satellites, for example).  In the search for point sources, a threshold of ten times the Root Mean Square (RMS) in the difference image was adopted and a Full Width at Half Maximum (FWHM) for the source of five pixels was adopted.

\begin{figure}[h!]
\begin{center}
\includegraphics[width=0.45\textwidth]{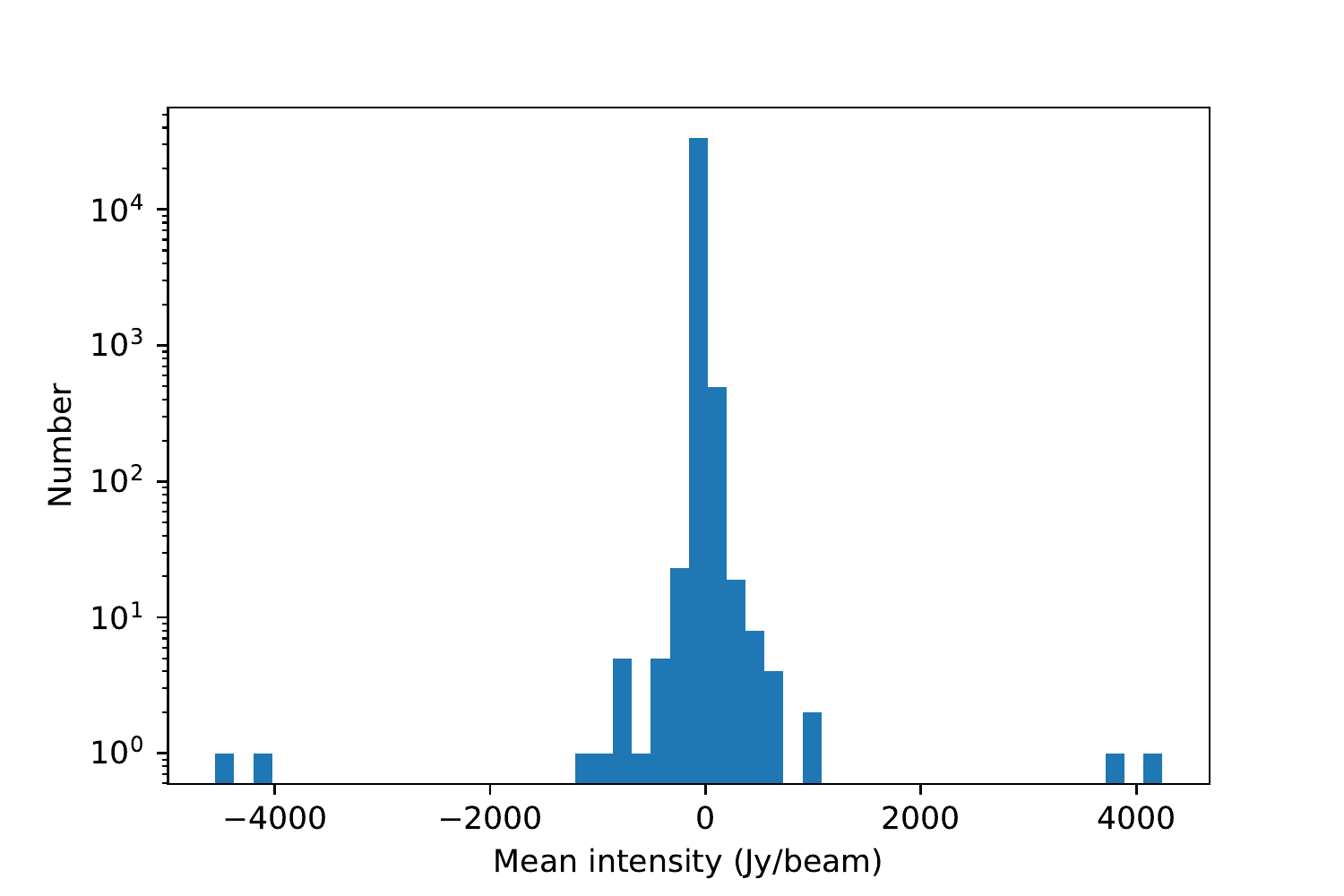}
\includegraphics[width=0.45\textwidth]{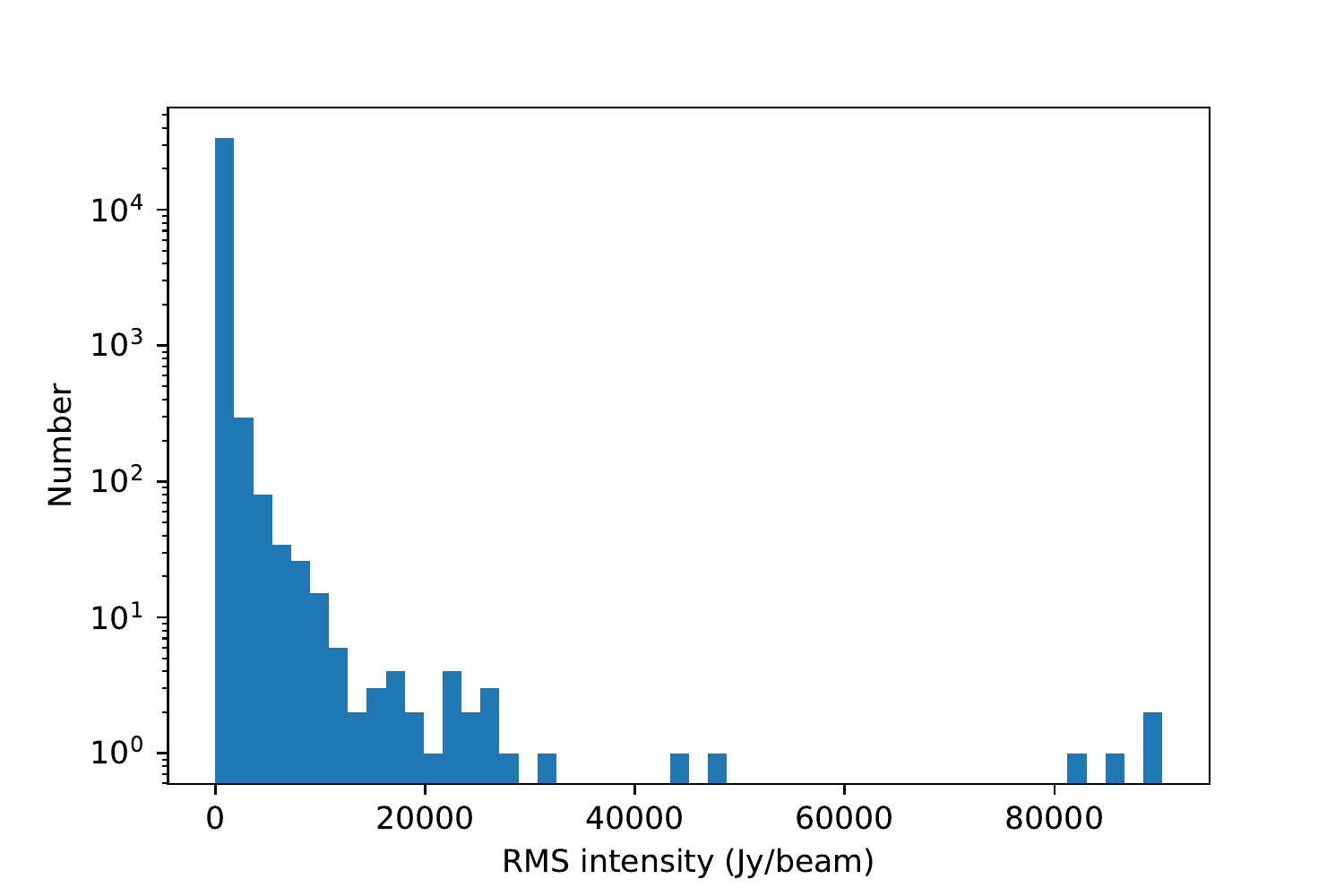}
\caption{Distributions of the mean of the difference images (top) and the RMS (bottom) for the XX polarisation data.}
\label{meanstd}
\end{center}
\end{figure}

The RMS values of the difference images vary.  For images with strong sidelobes due to sources of interference and/or the presence of strong astronomical sources, the RMS values can be higher than at other times.  Thus, the detection threshold of ten times the RMS varies across the set of observations.  Figure \ref{meanstd} shows the distributions of mean and RMS values, for all difference images in the XX polarisation.  The YY polarisation distributions are very similar and are not shown.

As can be seen in these figures, the RMS values are strongly dominated by values of less than $\sim$1000 Jy/beam, but with outliers of up to 100,000 Jy/beam.  Generally these outliers are encountered when very strong episodes of direct reception from distant transmitters (for example from Geraldton) occur, or when reflections from aircraft over the MRO are seen, or when a particularly strong meteor reflection is present.

The pixel coordinates, times, and peak intensities of all detections were recorded.  The detections were then visualised as density maps in pixel coordinates, celestial coordinates, and Galactic coordinates, in order to make an initial examination of the variety of signals present in the data.

Broadly speaking, the detected signals can be well categorised as the following: 

1) sources confirmed to a single time step occurring at random locations in the sky, with larger numbers occurring near the horizon ($<$20$^{\circ}$ elevation) than near the zenith (mostly likely meteor reflections); 

2) reflections from aircraft, identified as having very high peak intensities and being trackable from time step to time step, along known flight paths at angular speeds corresponding to standard aircraft air speeds and altitudes; 

3) candidate reflections from satellites, identified as being similar to planes but with far lower peak flux densities, detectable preferably near zenith, not having the trajectories of known flight paths, and with the expected angular speeds of objects in Low Earth Orbit; 

4) persistent sources of time-variable RFI on the horizon and at azimuths corresponding to population centres at large distances, beyond the MWA horizon; and

5) detections that followed the locations of the brightest celestial radio sources, largely due to imperfect difference imaging in the presence of the extremely bright radio sources, likely due to combinations of sidelobes around these sources, the effect of the ionosphere, and small errors in imaging and calibration.  

Across all categories, a total of 40,133 detections were made in the XX polarisation and a total of 39,954 in the YY polarisation.

This situation of a time variable RMS and a fixed signal-to-noise detection threshold somewhat problematic in terms of a robust statistical analysis of the full 67,988 image set and the different signal populations.

We could adopt a running median of the image RMS, to mitigate against the variations in RMS.  However, this would then effectively lower the signal-to-noise ratio of the detection threshold in a time variable manner.  For example, for an image with a high noise (due to sidelobes caused by the presence of a very strong transient signal) relative to the running median, and using a fixed detection threshold, the result would be a large number of false detections above the threshold for that image.  In this case, we would be corrupting the events we claim to be real.

The approach we have taken ensures that the events we claim to be real are real, at the price of occasionally sacrificing real events that would sit above the detection threshold for other images with lower noise levels.  Given the large volume of observations, and therefore total number of detections, we have clearly been able to distinguish between the different categories of signals, even though we have sacrificed some real signals.

We cannot claim to be able to do statistically robust population analyses for these signal types, in part due to the issue of the variable RMS and the loss of some real detections.  However, we maintain that even without this issue, such a detailed analysis would be difficult, due to: the calibration of the signal strengths because of primary beam effects; the intrinsic nature of the signals, which is highly dependent on the placement and illumination pattern of the RFI source (FM radio stations hundreds of km away); propagation effects; and the temporal and frequency resolution of our data.

\begin{figure}[h!]
\begin{center}
\includegraphics[width=0.45\textwidth]{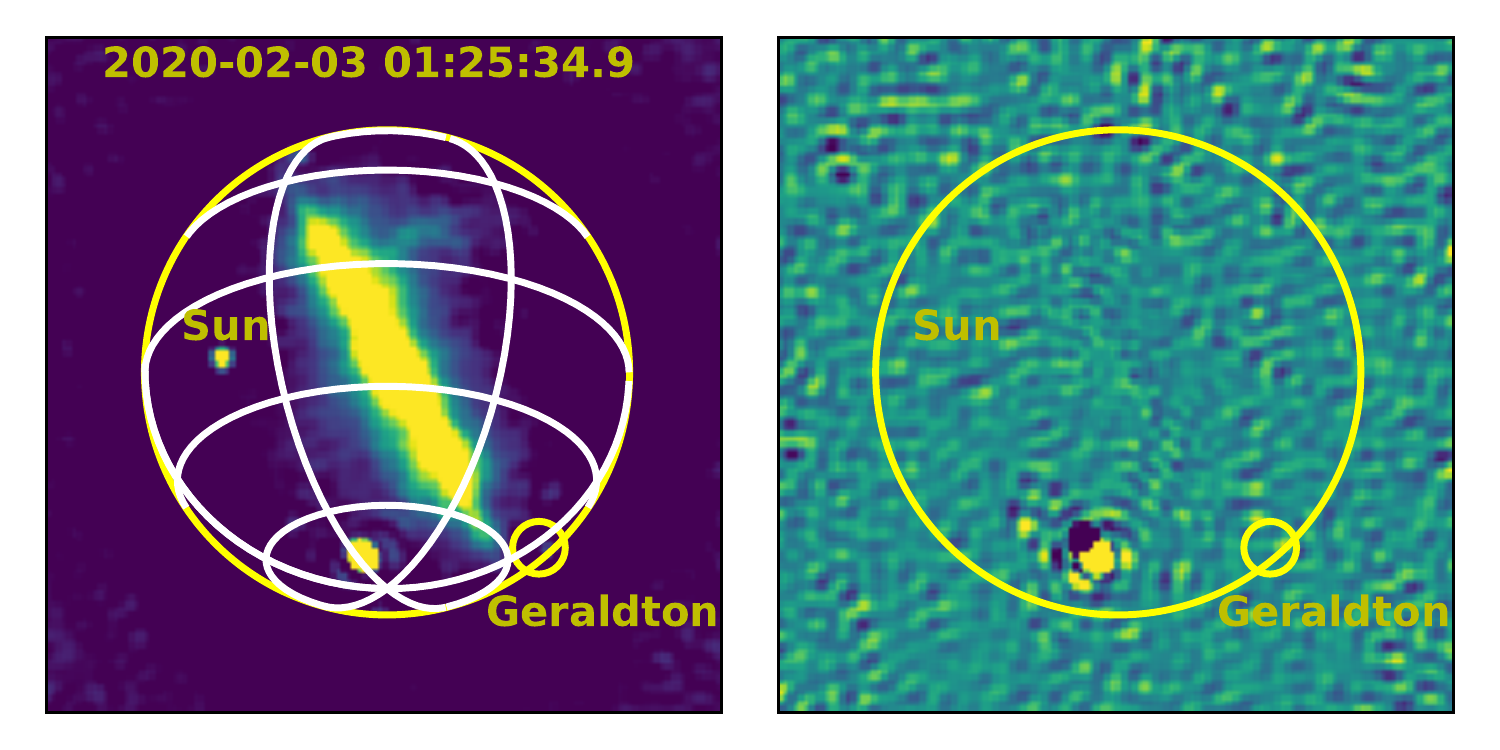}
\includegraphics[width=0.45\textwidth]{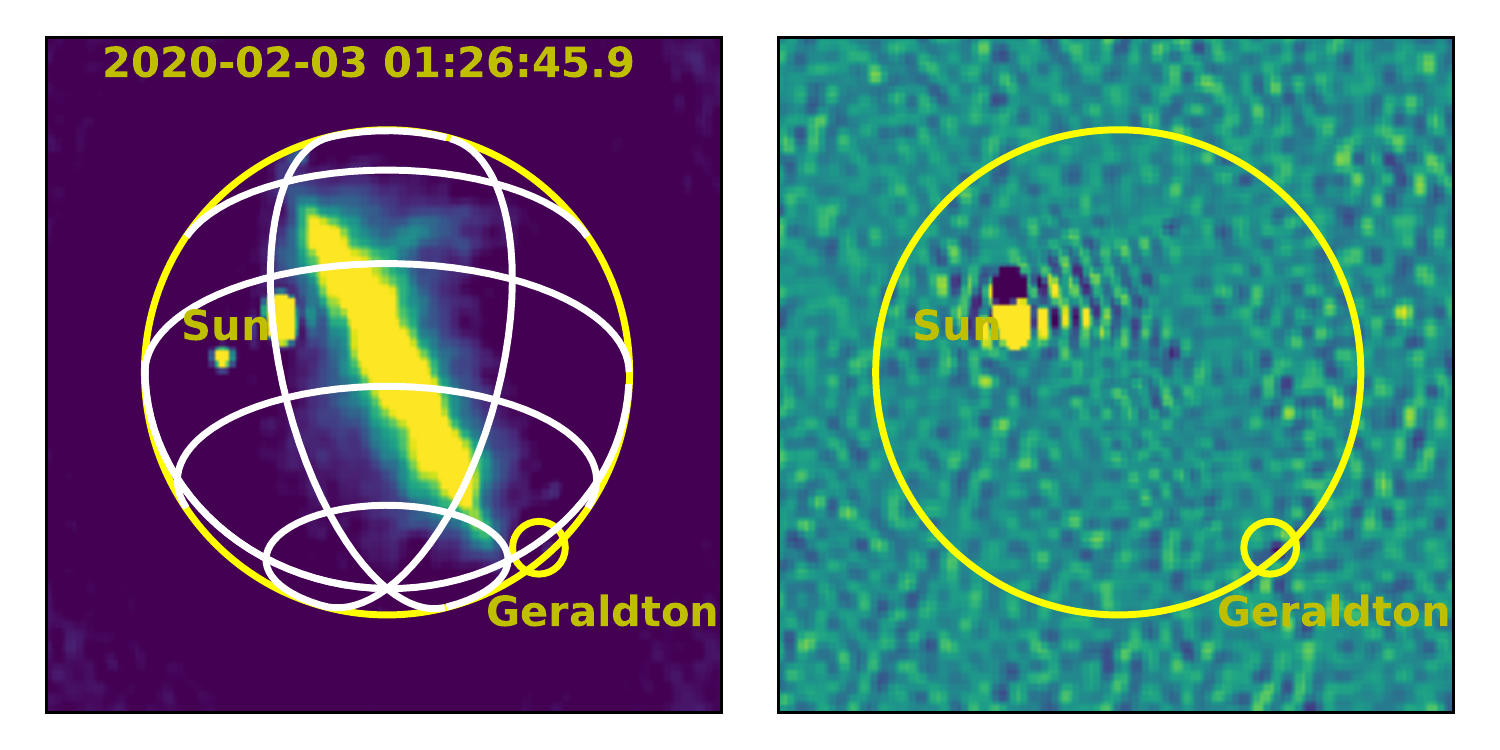}
\includegraphics[width=0.45\textwidth]{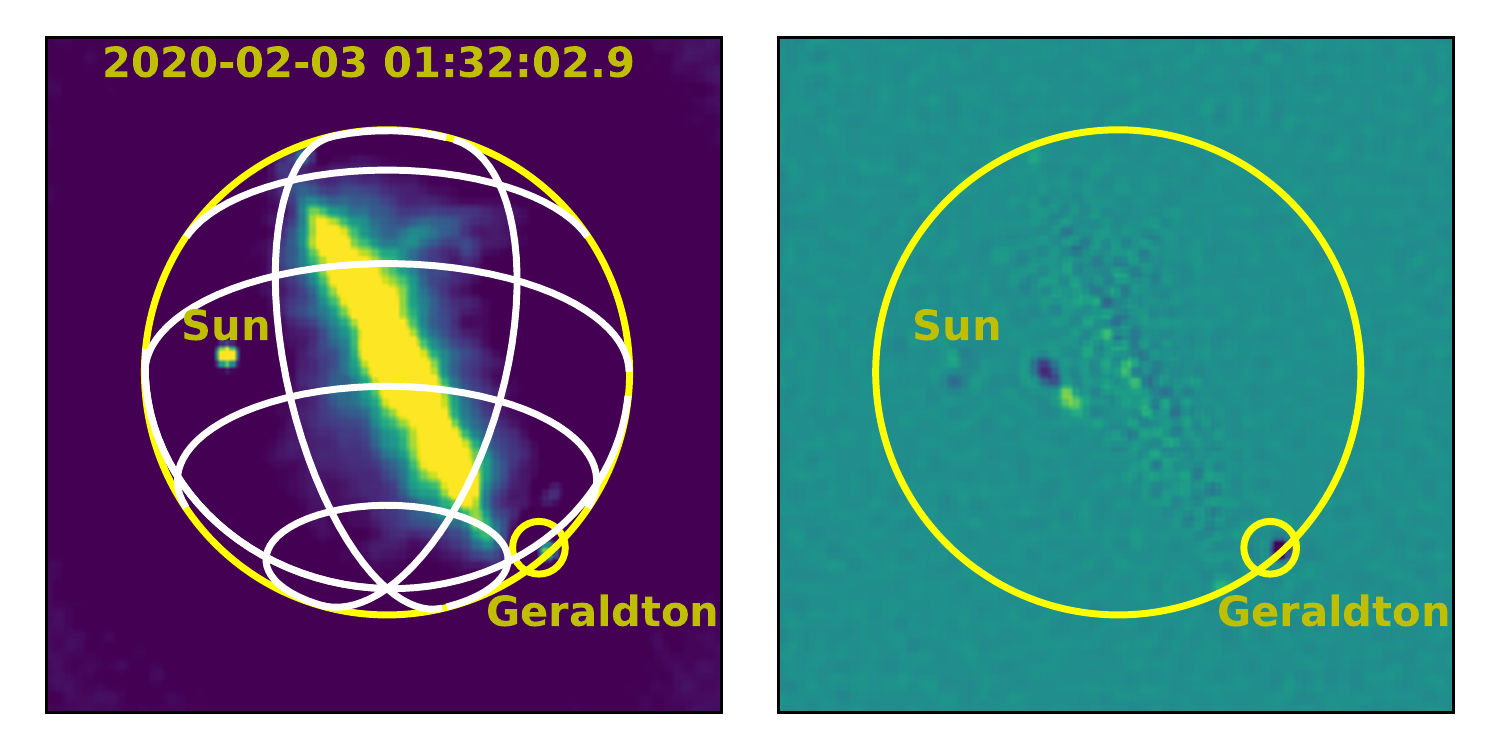}
\includegraphics[width=0.45\textwidth]{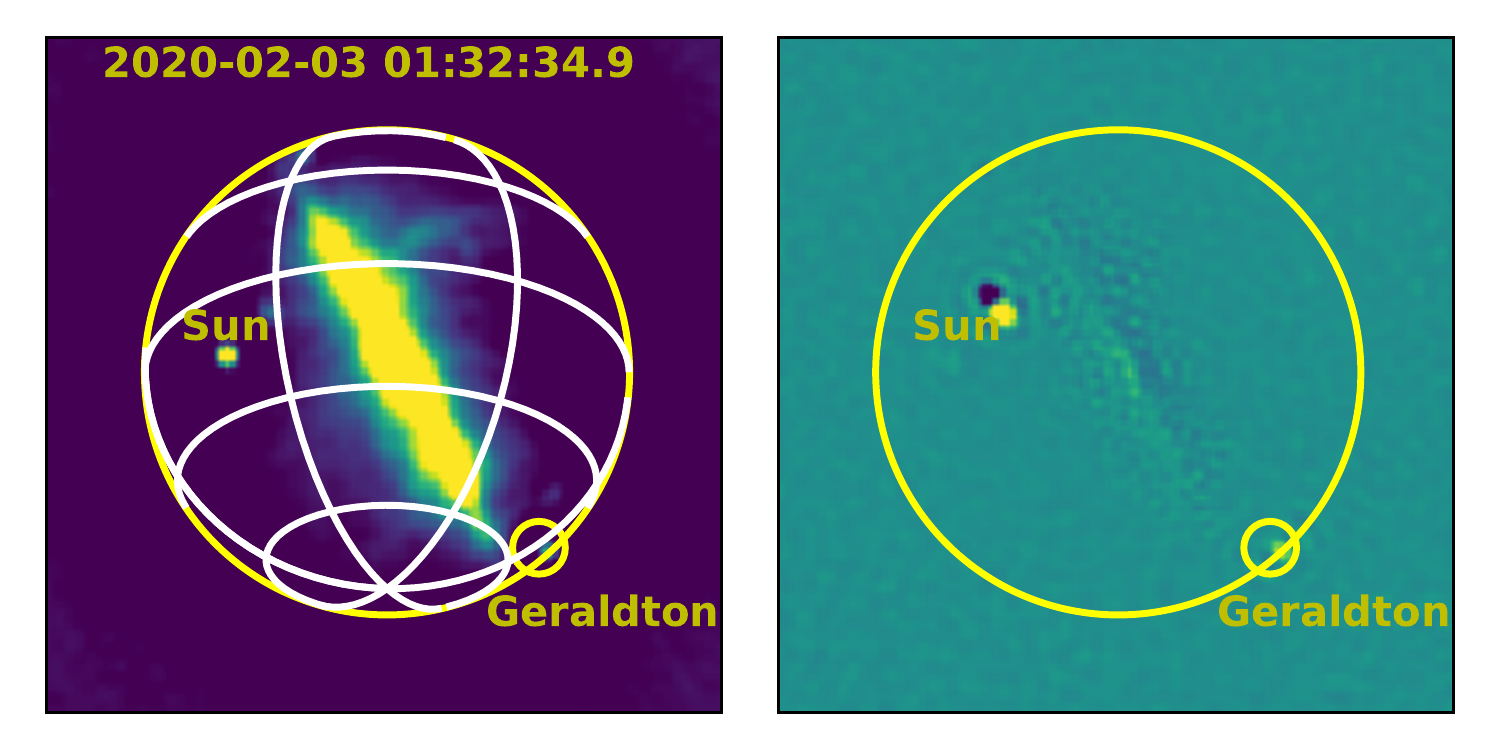}
\includegraphics[width=0.45\textwidth]{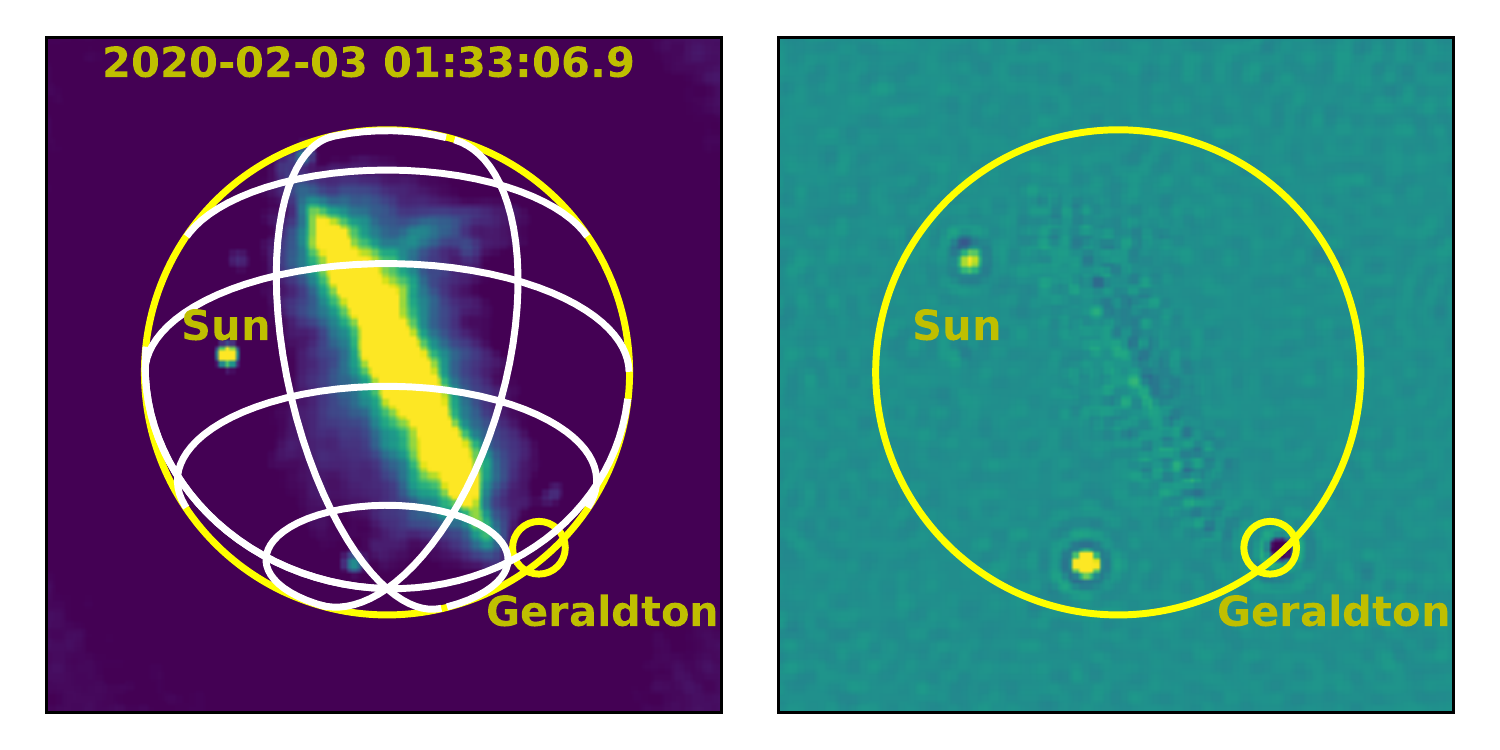}
\caption{A sequence of five images (left panels), plus corresponding difference images (right panels), for the XX polarisation, as described in the text (times are UTC).}
\label{image}
\end{center}
\end{figure}

Figure \ref{image} shows a sequence of five images (left panels; intensity range from 0 Jy/beam to 5000 Jy/beam), plus corresponding difference images (right panels; intensity range from -600 Jy/beam to 600 Jy/beam), with examples of the different categories of signal present in the EDA2 data, for the XX polarisation.  In all images, the Galactic Centre is near transit at the zenith.  The Sun is present and labeled and the location of Geraldton on the horizon is labeled.  From top to bottom (time order): the top two images show an aircraft moving from south to north, including high amplitude residuals due to sidelobes; the third, fourth, and fifth images show a satellite moving toward the north-east, between the Galactic Centre and the Sun.  For the aircraft and the satellite, the difference images show the distinctive streaks with positive/negative structures \citep{2020arXiv200201674P}.  The fifth image also shows a single timestep signal likely from a meteor, near the southern horizon.  Throughout the series of images, residuals near the Galactic Centre are apparent, also at the Sun (likely due to the strong variability of even the quiet Sun).  Variability of astronomical sources due to ionospheric scintillation will also make some contribution to difference image artifacts.  The variability of the signals received from over the horizon toward Geraldton is also apparent, in this case due to tropospheric scattering/ducting.

Movies showing the full sequences of the original images, for both XX and YY polarisations, are available as electronic supporting material in the online publication.

Further processing of the data was undertaken to isolate and examine the different categories of signal in more detail.  

For the category of signals associated with high amplitude residuals from the differencing process and associated with the brightest celestial radio sources in the sky, the simplest method to isolate them is to define exclusion regions around their locations.  Thus, for each difference image, a set of celestial radio source coordinates were transferred into image pixel coordinates using the WCS information attached to the images (utilising functions within the astropy Python module).  Exclusion regions with a radius of five pixels were defined around the sources and signals detected in these regions were excluded from further analysis.  The bright radio sources in question are: the Sun; Cygnus A; Centaurus A; 3C273; the Crab nebula; Hydra A; Fornax A; Virgo A; Pictor A; and the Vela nebula.  Additionally, the Galactic Centre is the brightest radio region in the sky and this region was also excluded, but in this case it was more convenient to define an exclusion region based on Galactic coordinates, such that $-15^{\circ}<b<15^{\circ}$ and $-25^{\circ}<l<25^{\circ}$ defined the region.

The simplest and most effective method to isolate signals from aircraft was to manually identify the time ranges during which they were visible to EDA2 and construct a data set containing aircraft and a data set with aircraft absent.  Given the strong scattered signals from aircraft often produced strong sidelobes, we did not attempt to recover other categories of sources during periods when aircraft were present.  In total, 9.82 hours of data included strong signals from aircraft, from a total of 75.35 hours of observations, corresponding to a temporal percentage of 13\%.  \citet{https://doi.org/10.7907/25dp-j474} accurately describes aircraft as ``an irritating source of RFI''.  Details of the characteristics of the aircraft signals are given in \S \ref{subsec:air}.

A similar approach was taken for the isolation of satellite signals.  In total, seven candidate satellite signals were detected, occupying approximately 16 minutes of the total observation time, approximately 0.3\%.  These candidates were confirmed via examination of the original images and were verified by comparing the observed trajectories to trajectories predicted from orbital parameters corresponding to the epoch of observation.  In all seven cases, a positive identification with an object in orbit could be made.  Detailed information on these satellite detections is given in \S \ref{subsec:orb}.  Although the fraction of the observations containing satellites is small, we exclude these times from the analysis of other categories of source in the same way that we exclude aircraft.

The isolation of signals from persistent transmitters beyond the MWA horizon is generally relatively straightforward, since the signals occupy a constant position in the images.  However, because the difference images will only characterise the RFI variability between time steps of a persistent transmitter, and because we are interested in the apparent intensity of the RFI, we extract signals from the original images and describe the signals in detail in \S \ref{subsec:tran}.  In terms of excluding these signals from the difference image analysis, we place exclusion regions around the transmitter locations most affected.  The strongest RFI originates from Geraldton, south-west of the MRO.  For this transmitter origin, a circular region of exclusion in the difference images corresponding to a radius of ten pixels was used.  Signals from this region were not considered further in the difference image analyses.  Likewise, exclusion regions with a radius of ten pixels were located at the horizon positions of Karatha and Port Hedland, both due north of the MRO.

While transmitters on the horizon were generally straightforward to deal with, we did observe some extreme variability on the horizon (almost the entire western horizon) in the last $\sim$1.8 hours of the observation period.  This behaviour was not localised and was extreme in its intensity and positional variability, particularly in the YY polarisation data (more sensitive in the east-west direction than the XX polarisation).  This activity is due to lightning and we describe it in detail in \S \ref{subsec:tran}.  Thus, the last 1.8 hours of data were also excluded from further analysis of the difference images.

After the isolation of residual errors near bright radio sources, aircraft, satellites, persistent signals on the horizon from transmitters, and the extreme activity noted in the final 1.8 hours of the observation period, the remaining signals in the difference images represent signals confined to single time steps and occurring randomly across the sky.  Overwhelmingly, these signals are likely to be due to backscatter from meteor trails in the upper atmosphere.  Detailed information for these signals is presented in \S \ref{subsec:met}.  15,530 of these signals were detected, in both XX and YY polarisations, representing approximately 37\% of the total number of detected signals.

\section{RESULTS}
\label{sec:res}

For each category of signal we have described and isolated in the previous section, we provide detailed information below.  In general, we examine the spatial, temporal, and intensity distributions of these signals and show examples of individual signals.  Throughout this section, many of the figures depicting histograms of peak intensity for the various classes of event appear to have powerlaw-like distributions.  In all cases, we attempted to parameterise the distributions with powerlaws, but in no case did a single powerlaw provide a good description of any observed distribution.

\subsection{Aircraft backscatter}
\label{subsec:air}

\begin{figure}[h!]
\begin{center}
\includegraphics[width=0.45\textwidth]{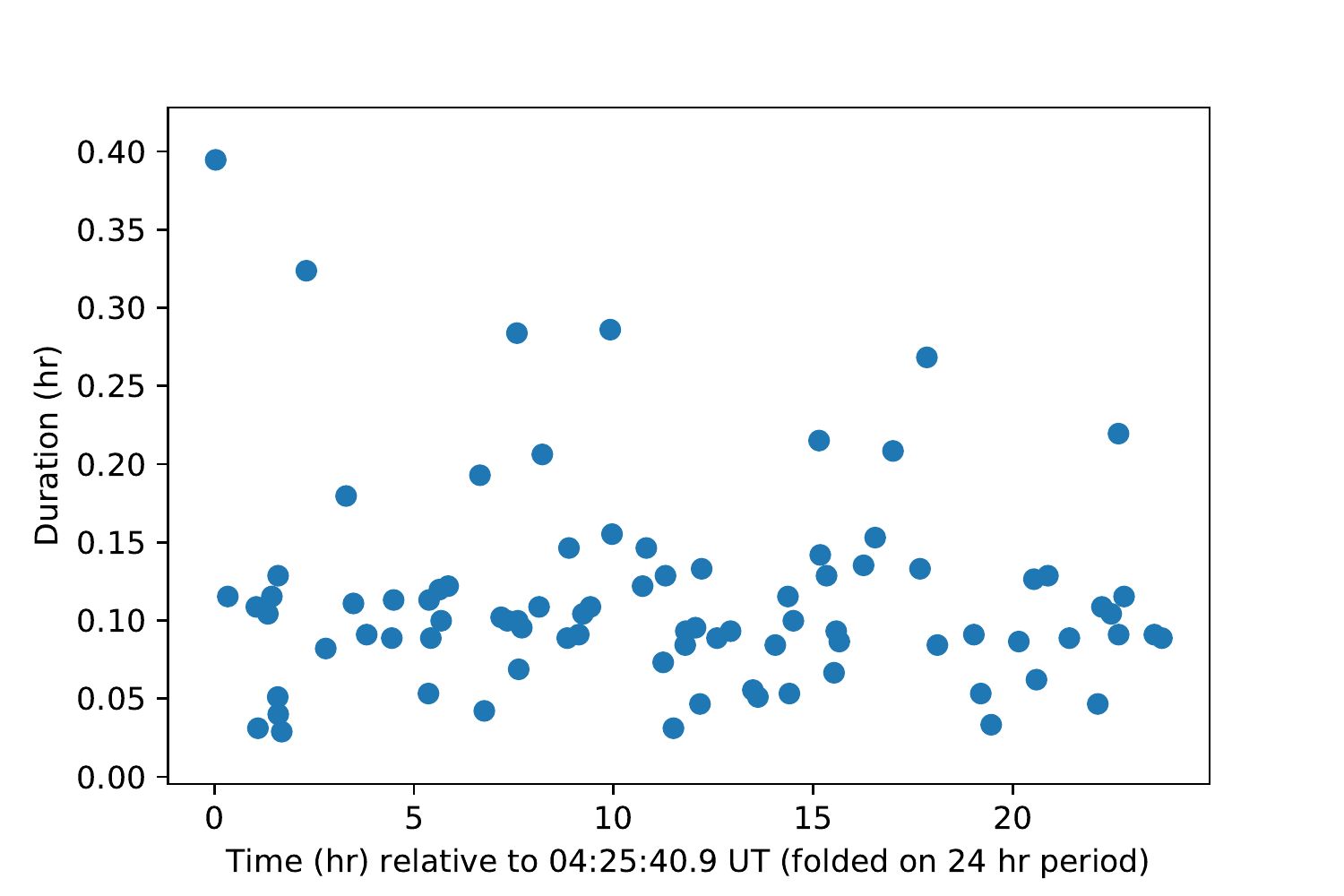}
\caption{Durations of periods in which aircraft backscatter signals are present in the EDA2 images, as a function of time relative to the starting time of observations (and folded on a 24 hour period, given that flight schedules from day to day are similar).}
\label{pdur}
\end{center}
\end{figure}

Figure \ref{pdur} shows the durations of periods in which aircraft are visible in the EDA2 data, as a function of time of day.  Each point in Figure \ref{pdur} denotes a discrete period during which one or multiple aircraft are present.

\begin{figure}[h!]
\begin{center}
\includegraphics[width=0.4\textwidth]{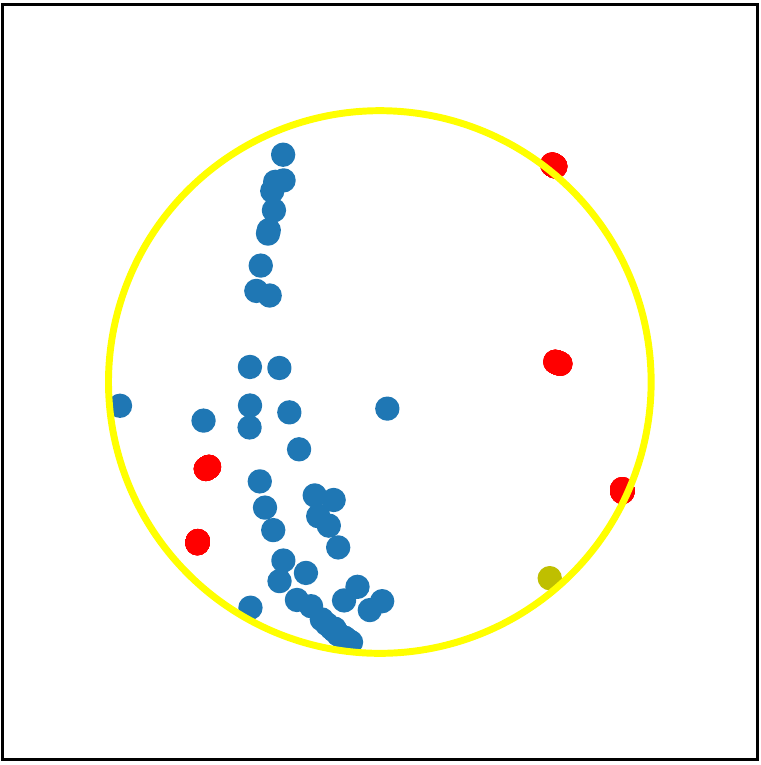}
\includegraphics[width=0.4\textwidth]{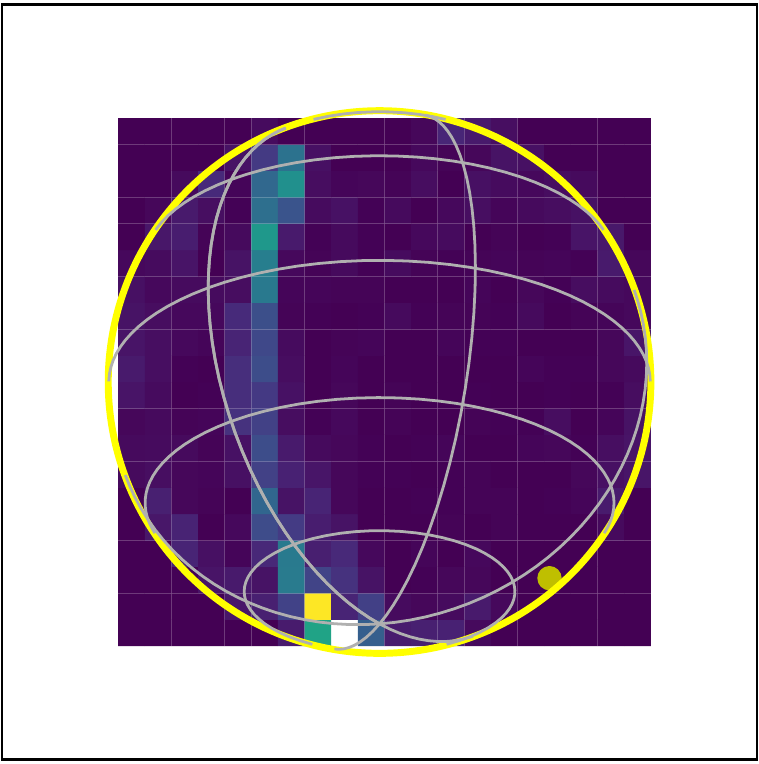}
\caption{Example of two aircraft trajectories across the sky on north-south flight paths (top panel) and the two dimensional histogram of detected signals during the periods when aircraft are visible from the MRO (bottom panel).  The red markers in the top panel denote the positions of strong astrophysical radio sources above the horizon at this time.  In both panels, the yellow marker denotes the location of Geraldton on the horizon.}
\label{pexa}
\end{center}
\end{figure}

Figure \ref{pexa} (top panel) shows example aircraft trajectories, as detected in the XX polarisation via the difference imaging analysis, over an approximate 6.5 minute period.  As noted in \S \ref{subsec:ann}, the integrated observation time affected by RFI reflected from aircraft represented in Figure \ref{pdur} is 9.82 hours, or 13\% of the total observation time over three days.  Over a multi-day period, this will be representative of the general situation for the MRO.  Overwhelmingly, the flight paths over the MRO are north-south routes, to/from Perth, which is due south of the MRO, to/from a number of mining centres in the north of Western Australia, and to/from a range of international locations largely north of Australia.  A small number of east-west routes to the north of the MRO are also detected, and also other seldom-used flight paths.

A two-dimensional histogram of the spatial occurance of signals detected during the 9.82 hour period is also illustrated in Figure \ref{pexa} (bottom panel), reflecting the preponderance of north-south flight paths over the MRO.

\begin{figure}[h!]
\begin{center}
\includegraphics[width=0.45\textwidth]{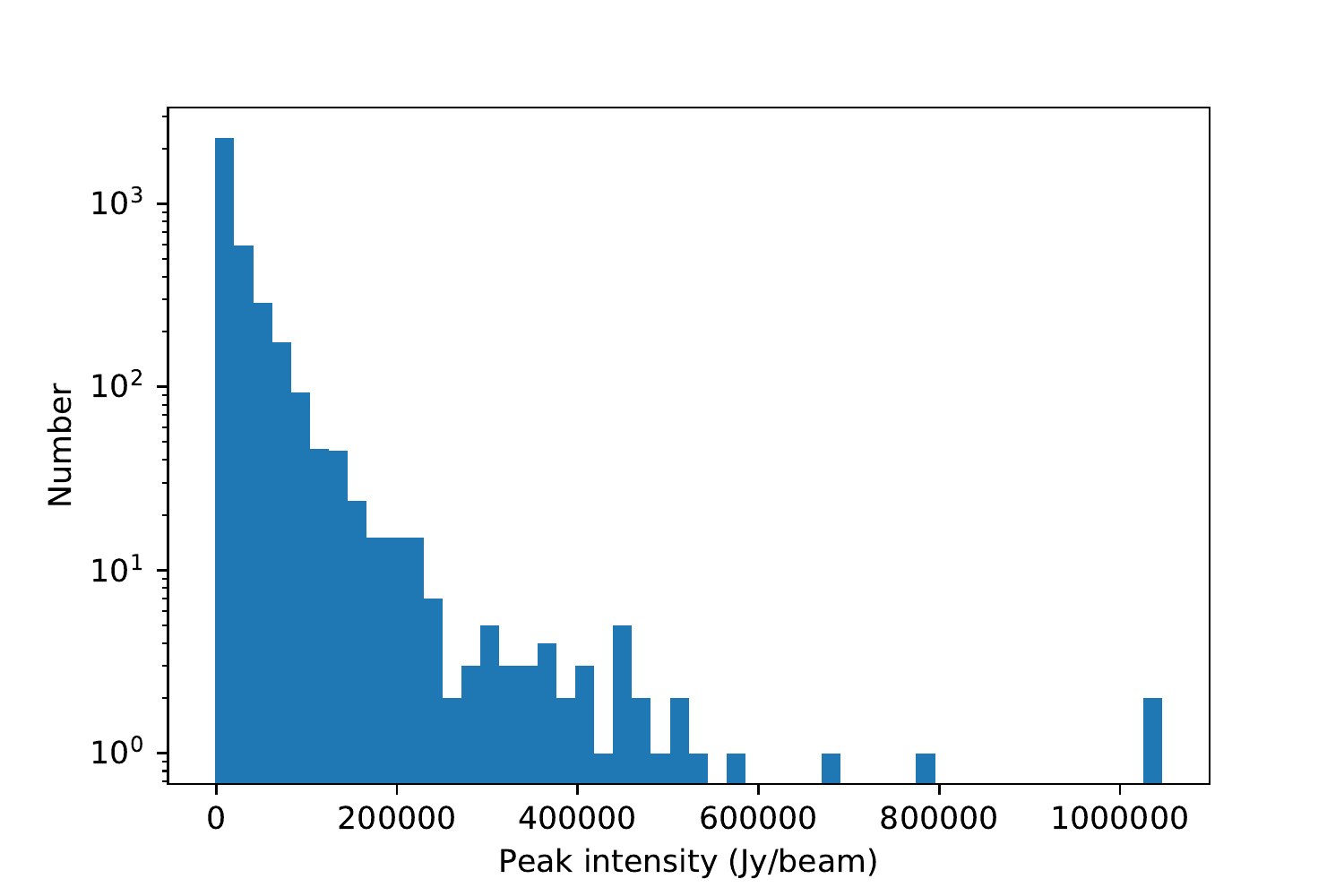}
\caption{Peak intensity of signals detected during periods in which aircraft are above the horizon at the MRO, for the XX polarisation.}
\label{plaint}
\end{center}
\end{figure}

Finally, Figure \ref{plaint} shows the distribution of peak intensity of signals in the XX polarisation (the distribution for the YY polarisation is effectively identical) for periods during which aircraft are visible in the EDA2 data.  The signals shown in Figures \ref{pexa} and \ref{plaint} account for detections of aircraft, but also include all other detections of other categories of signals present during these periods.  The detections are dominated by aircraft detections, however, characterised by their very high peak intensities as compared to other signal categories, as shown in later sections.

\subsection{In-orbit object backscatter/transmission}
\label{subsec:orb}

Artificial satellites in Earth orbit also produce backscattered signals from FM transmitters, in the same way that aircraft do.  Even though satellite passes are vastly more frequent above the MRO horizon (constant, in fact) than aircraft, the objects are much more distant ($\sim$1000 km vs $\sim$10 km) and generally smaller ($\sim$1 m vs $\sim$100 m), and thus produce far weaker signals at the MRO.  

However, as demonstrated using the MWA, a sensitive instrument at FM frequencies can detect many reflections from satellites \citep{2020arXiv200303947H,2020arXiv200201674P,8835821,7944483,Tingay2013OnFeasibility} at high signal to noise.  Although the EDA2 instrument is not as sensitive as the MWA (approximately 1/10$^{th}$ as sensitive), a handful of satellite detections have been made.

Table \ref{satdet} provides the list of satellites identified from the difference imaging process, based on being able to track a trajectory across the sky over multiple time steps.

\begin{table*}[h!]
\caption{Identified backscatter from objects in orbit and their properties}
\begin{tabular}{|c|c|c|c|c|c|} \hline \hline
Satellite$^{a}$   &NORAD           &Start      &End               &Mean intensity         &RCS$^{b}$            \\
name        &ID \#           &time (UT)       &time (UT)             & (Jy/beam)      &($m^{2}$)        \\ \hline
BGUSAT      & 41999 &2020-01-31 14:40:09.9  &2020-01-31 14:43:11.9  &1060  &$<$0.1       \\ \hline
ISS (ZARYA) & 25544 &2020-01-31 17:17:41.9  &2020-01-31 17:19:00.9  &440  & $>$1.0      \\ \hline
MAX VALIER SAT& 42778 &2020-02-03 01:18:16.9  &2020-02-03 01:20:09.9  &650  &0.1 $-$ 1.0       \\ \hline
ISS (ZARYA) &22554 &2020-02-03 01:31:07.9  &2020-02-03 01:33:14.9  &930  &$>$1.0         \\ \hline 
FLOCK 3P 71 & 42024 &2020-02-01 14:16:22.9& 2020-02-01 14:18:05.9  &330  &$<$0.1       \\ \hline   
ISS (ZARYA) & 25544 &2020-02-02 02:18:17.9  &2020-02-02 02:20:16.9  &740  &$>$1.0 \\ \hline 
BGUSAT      & 41999 &2020-02-02 02:18:17.9  &2020-02-02 02:20:16.9  &1010  &$<$0.1 \\ \hline \hline
\end{tabular}\\
$^{a}$ TLE information for predicted trajectories from space-track.org for the epoch of observation: 2020-02-01 \\
$^{b}$ Radar Cross Section (RCS) is categorised by space-track.org as: small ($<$0.1); medium (0.1 $<$ RCS $<$ 1.0); and large ($>$1.0) \\
\label{satdet}
\end{table*}

\begin{figure}[h!]
\begin{center}
\includegraphics[width=0.4\textwidth]{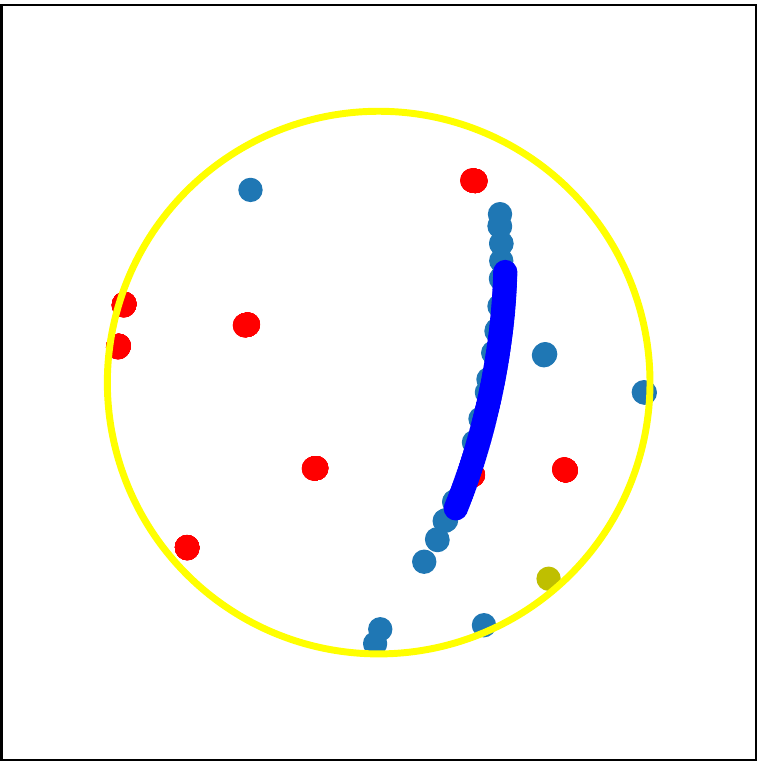}
\caption{Example satellite trajectory detected as described in the text, for the BGUSAT satellite as listed in Table \ref{satdet}, compared to the trajectory predicted from orbital elements at the epoch of observation.  Only the middle portion of the predicted trajectory is shown as dark blue markers, to allow comparison to the observed trajectory (light blue markers).}
\label{sate}
\end{center}
\end{figure}

Figure \ref{sate} shows an example of a detected satellite trajectory, for the BGUSAT satellite, compared to the trajectory predicted using orbital parameters from the epoch of observation (utilising the PyEphem package in python).

Table \ref{satdet} shows a number of detections of the International Space Station (ISS), which is not surprising given that this is the largest artificial satellite in Earth orbit and has been detected many times in Space Situational Awareness (SSA) passive radar tests and observations with the MWA \citep{Tingay2013OnFeasibility, 7944483, 2020arXiv200201674P}.  

More surprising are the detections of small satellites such as BGUCAT, MAX VALIER SAT, and FLOCK 3P 71.  BGUSAT is a nanosatellite developed at the Ben-Guiron University in Israel and was launched in February 2017; it is equipped with cameras for climate research\footnote{https://in.bgu.ac.il/en/epif/Pages/BGU-SAT.aspx}.  MAX VALIER SAT is a small Italian/German amateur satellite, launched in June 2017; it is equipped with an X-ray telescope\footnote{http://www.maxvaliersat.it/}.  FLOCK 3P 71 is a 3U cubesat that is part of a large fleet of satellites equipped with cameras for Earth observation; it was launched in February 2017\footnote{https://space.skyrocket.de/doc\_sdat/flock-1.htm}.

Given the low Radar Cross Sections (RCSs) for these three objects, we would not typically expect to detect them in reflected FM transmissions.  Some small satellites carry long antenna systems for communications at frequencies near 150 MHz, which may increase the RCS at FM frequencies.  However, even in these cases we would expect detection to be marginal.  In Table \ref{satdet}, BGUSAT is the brightest object detected, brighter than the ISS, which would not be expected (note however that these measured intensities are not corrected for the antenna beam pattern in Table \ref{satdet}.

It is possible that we are not seeing these small satellites in reflected FM, but are seeing direct transmissions from the satallites in the EDA2's frequency range.  Using the MWA for non-coherent passive radar observations, \citet{2020arXiv200201674P} found two cubesats to be transmitting broadband signals.  One was the Israeli student satellite Duchifat-1 and the other was the UK satellite UKube-1.  Indeed, there is evidence that BGUSAT is in the same situation; EDA2 data similar to those presented here, but at a frequency of 70 MHz (well below the FM band), shows a detected signal from BGUSAT (X. Zhang, Private Communication).  An investigation of MAX VELIER SAT and FLOCK 3P 71 should be conducted at other frequencies, to determine if they are also generating broadband transmissions.  These satellites are not licensed for broadband transmissions, which raises a potentially concerning situation for ground-based radio astronomy.

\begin{figure}[h!]
\begin{center}
\includegraphics[width=0.45\textwidth]{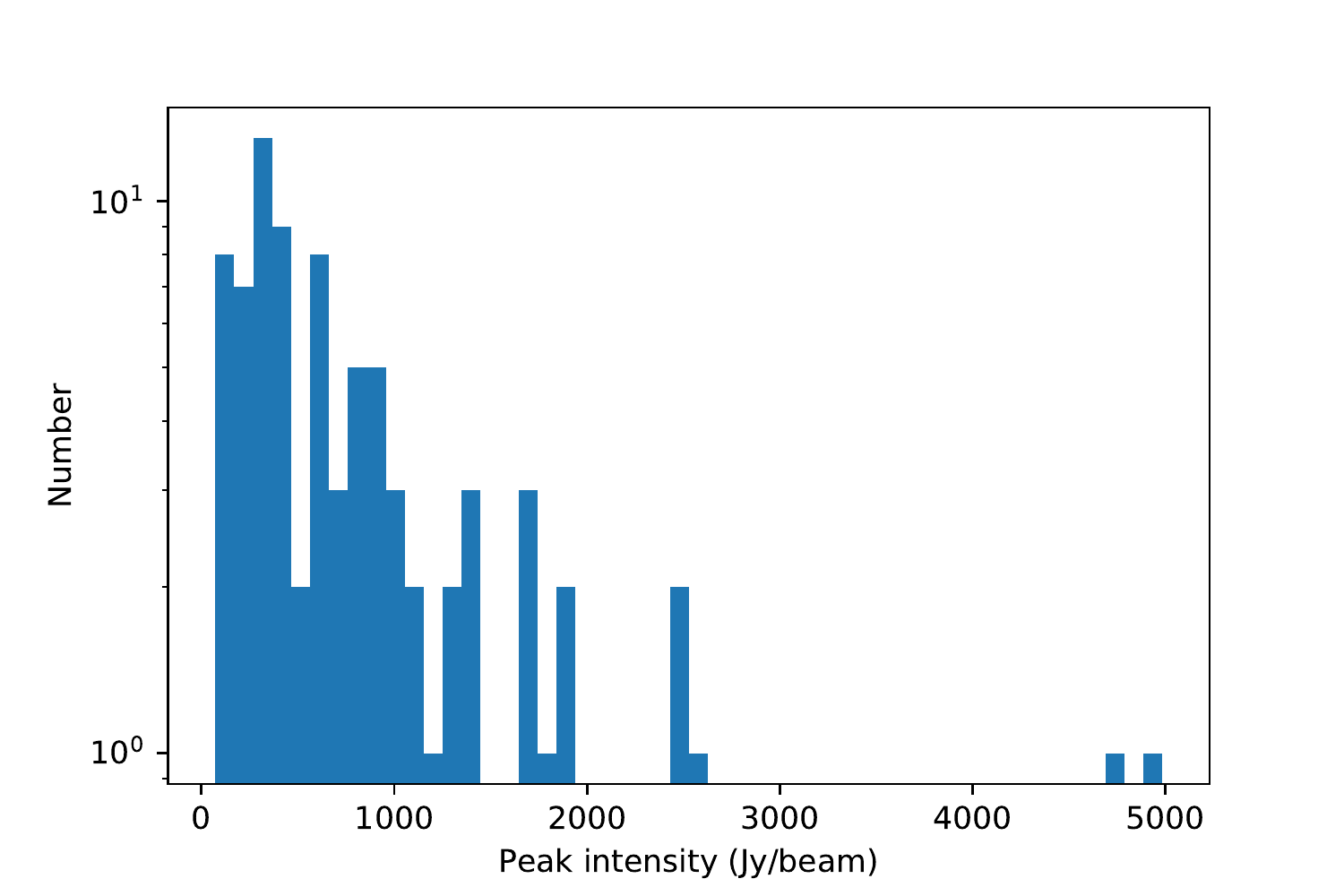}
\caption{Peak intensity of signals detected during periods in which satellites are detected in the EDA2 data (reflections and/or direct transmissions, as discussed in the text).}
\label{satfd}
\end{center}
\end{figure}

Finally, Figure \ref{satfd} shows the distribution of peak intensity of signals in the XX polarisation (the distribution for the YY polarisation is virtually identical) for periods during which satellites are detected in the EDA2 data (note that values much higher than the mean peak intensity listed in Table \ref{satdet} are seen in Figure \ref{satfd}).

\subsection{Transmitters on the horizon}
\label{subsec:tran}
As opposed to reflected signals from aircraft and satellites, the (more-or-less) direct reception of signals from FM transmitters is also strongly present in our data.  Although these transmitters do not have a line-of-sight to the EDA2, the signals can be propagated long distances via tropospheric scattering, and occasionally very strongly via tropospheric ducting events \citep{2017rfi..confE...1S,2016arXiv161004696S}.  FM transmitters are also present in proximity to virtually all towns and settlements across Western Australia, so the reception at EDA2 is not unexpected.  In particular, we persistently detect highly variable signals from the directions of Geraldton (south-west of the MRO), Karatha (north of the MRO), and Port Headland (north of the MRO).  Evidence for similar signals toward other directions, at much lower levels were noted in the data but are not expanded upon here.  We present some examples of the types of behaviours seen toward Geraldton, as the closest location with the strongest transmitters at this particular FM frequency.

In order to illustrate these signals, we take two approaches.  First, because these signals are persistent at a fixed location in our images, a difference image approach will only detect the variation of the signal with time.  More useful is to understand the fixed and variable components of the flux density, extracting these measurements from the original images rather than from the difference images.  However, we need to accept in this case that these measurements will also pick up any astronomical contribution at this image location as strong astronomical sources set in the direction of Geraldton.

\begin{figure}[h!]
\begin{center}
\includegraphics[width=0.45\textwidth]{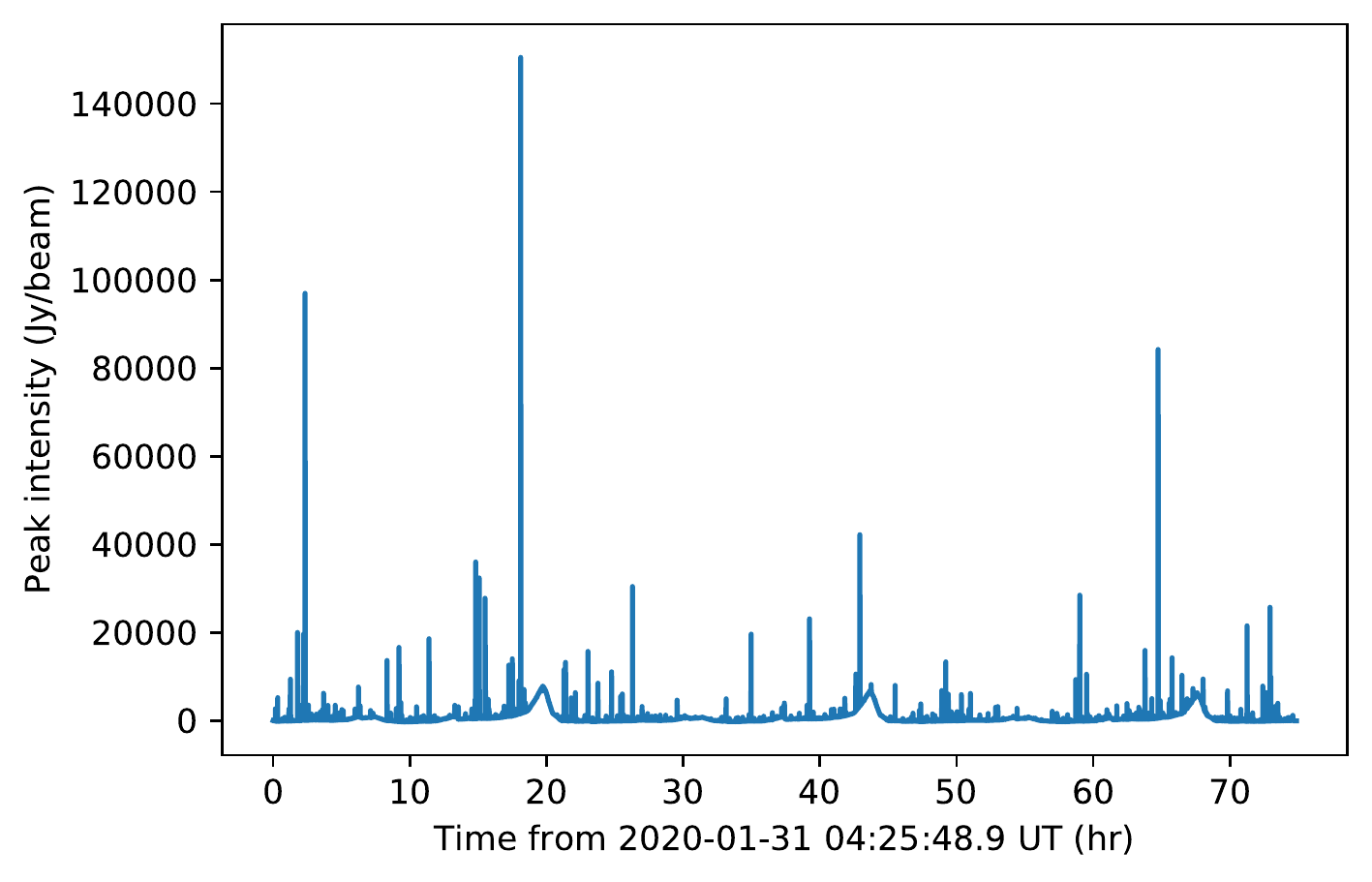}
\includegraphics[width=0.45\textwidth]{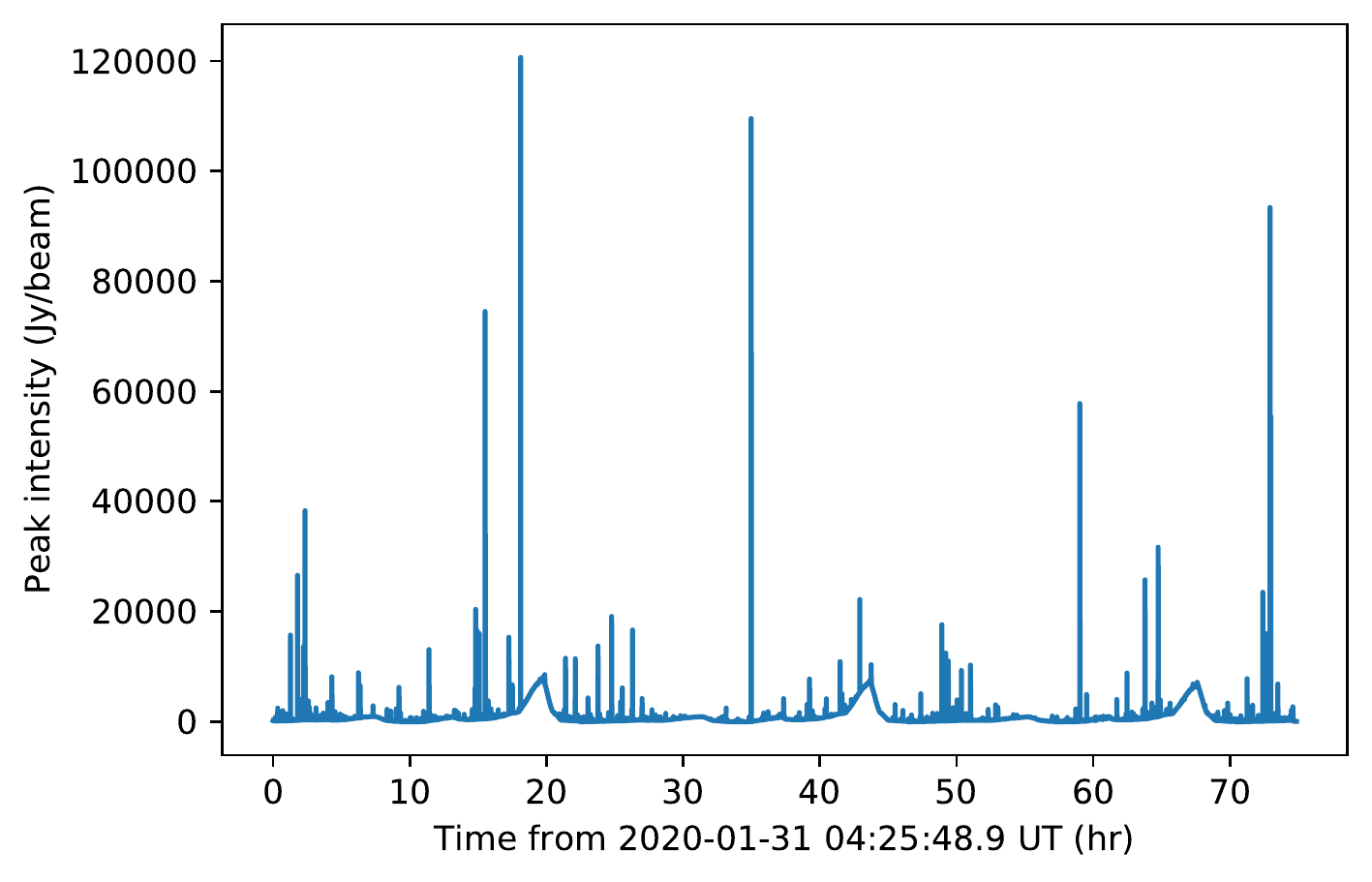}
\caption{The intensity as a function of time at the horizon in the direction of Geraldton, obtained from the original images for XX polarision (top) and YY polarisation (bottom) data.}
\label{ger}
\end{center}
\end{figure}

In Figure \ref{ger}, we show the intensity at the horizon in the direction of Geraldton, as a function of time, for both XX and YY polarisation data.  As can be seen, the signal is persistent but highly variable.  The baseline intensity also clearly shows the diurnal signal due to astronomical sources, repeating three times over the three days.  Particularly prominent is the setting of the Galaxy in the direction of Geraldton.

\begin{figure}[h!]
\begin{center}
\includegraphics[width=0.45\textwidth]{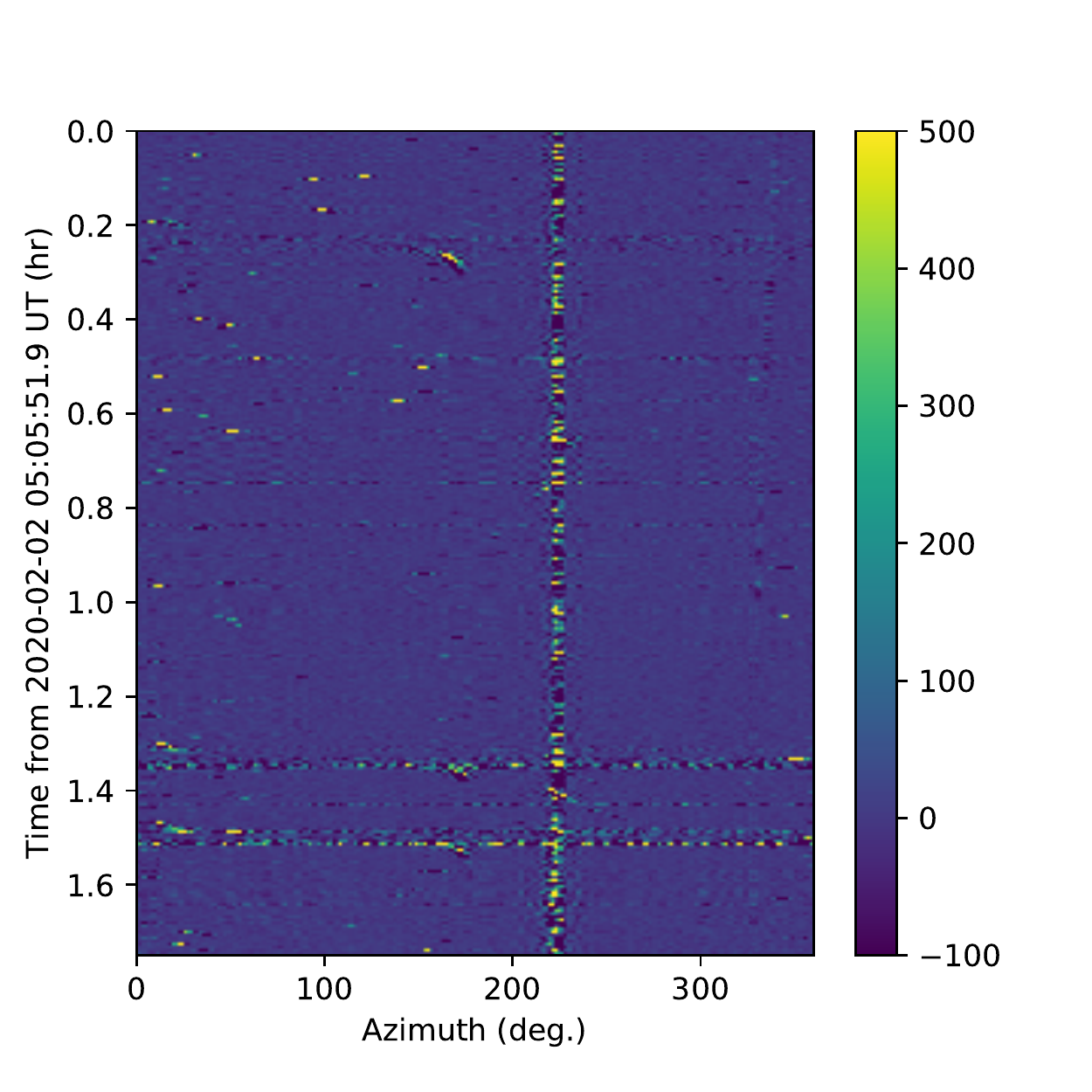}
\includegraphics[width=0.45\textwidth]{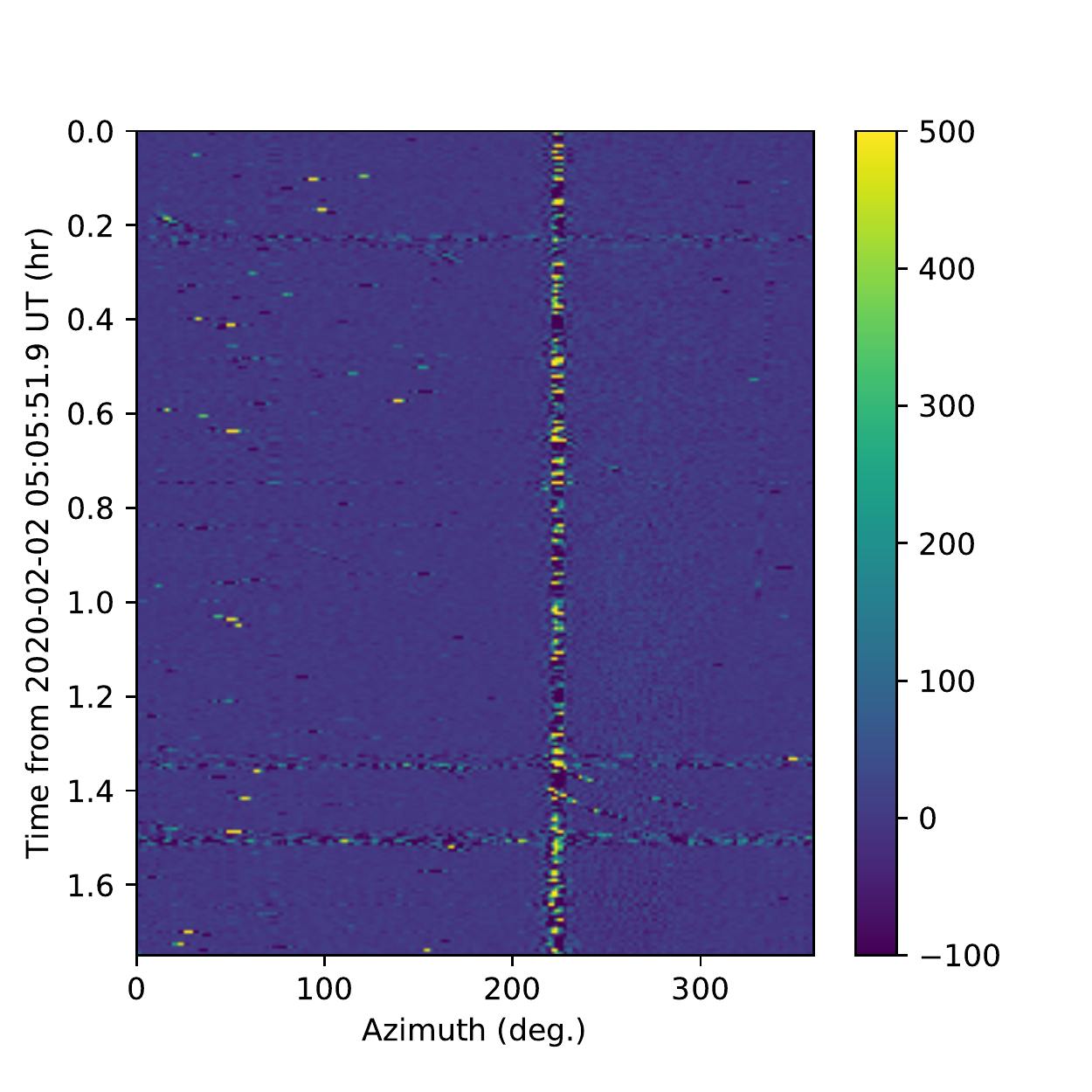}
\caption{The horizon intensity (colourbar units of Jy/beam) extracted from the difference images for a 1.8 hour period on the second day of observations, in XX (top) and YY( bottom) polarisations, as a function of azimuth angle (horizontal axis) and time (vertical axis).}
\label{hoxp}
\end{center}
\end{figure}

Second, we examine the difference image data, but across the full 360$^{\circ}$ horizon, in order to highlight some unusual behaviour noted earlier in the paper during the last $\sim$1.8 hours of the observations.  In Figure \ref{hoxp}, we show the intensities extracted from the difference images for an approximate 1.8 hour period during the second of the three days of observations, for both XX and YY polarisations.  The intensities are plotted as a function of azimuth angle (east of north) and time.  Clearly seen in both polarisations is the persistent and variable transmitter in Geraldton, with little activity at other azimuth angles.  This is typical of the behaviour throughout the observations.

\begin{figure}[h!]
\begin{center}
\includegraphics[width=0.45\textwidth]{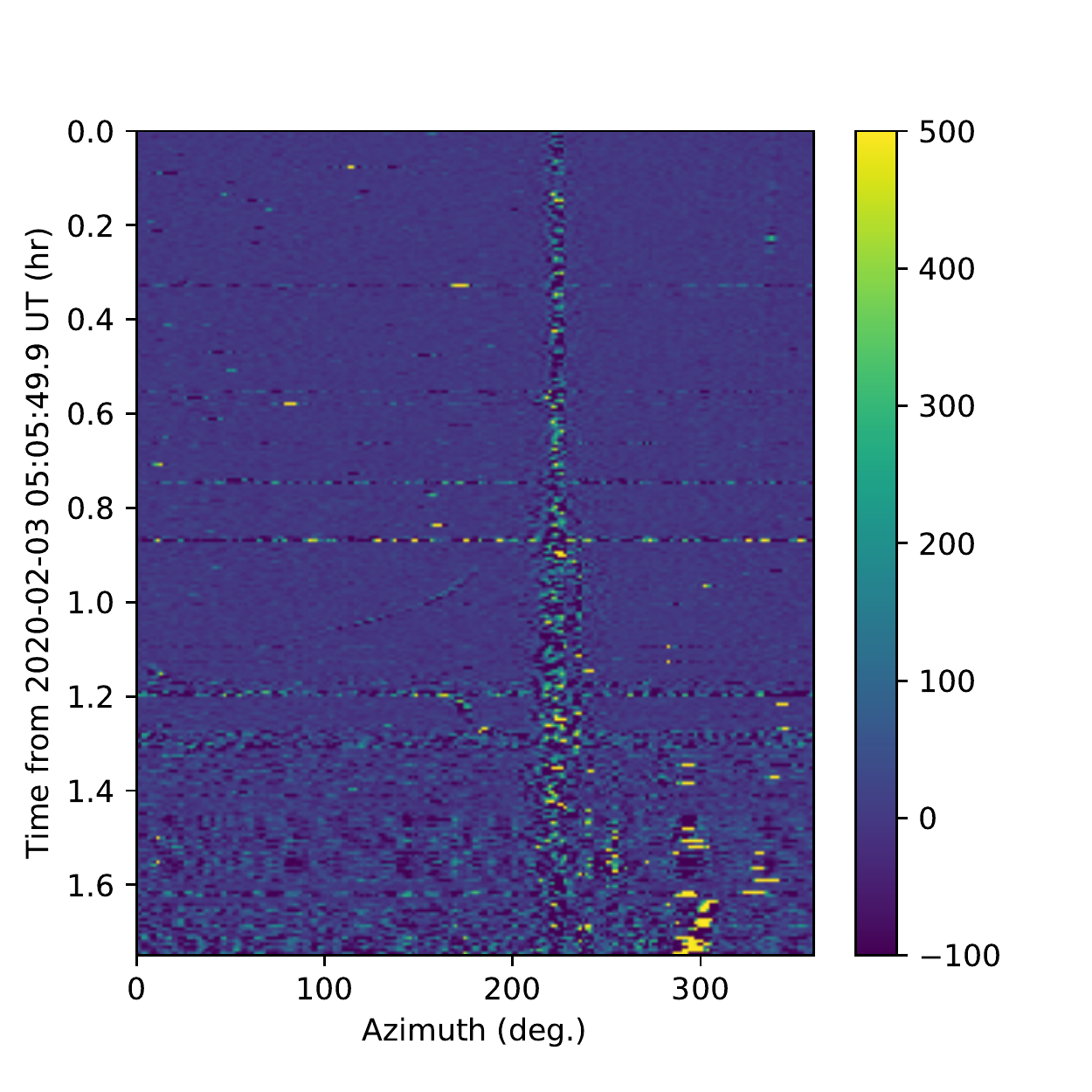}
\includegraphics[width=0.45\textwidth]{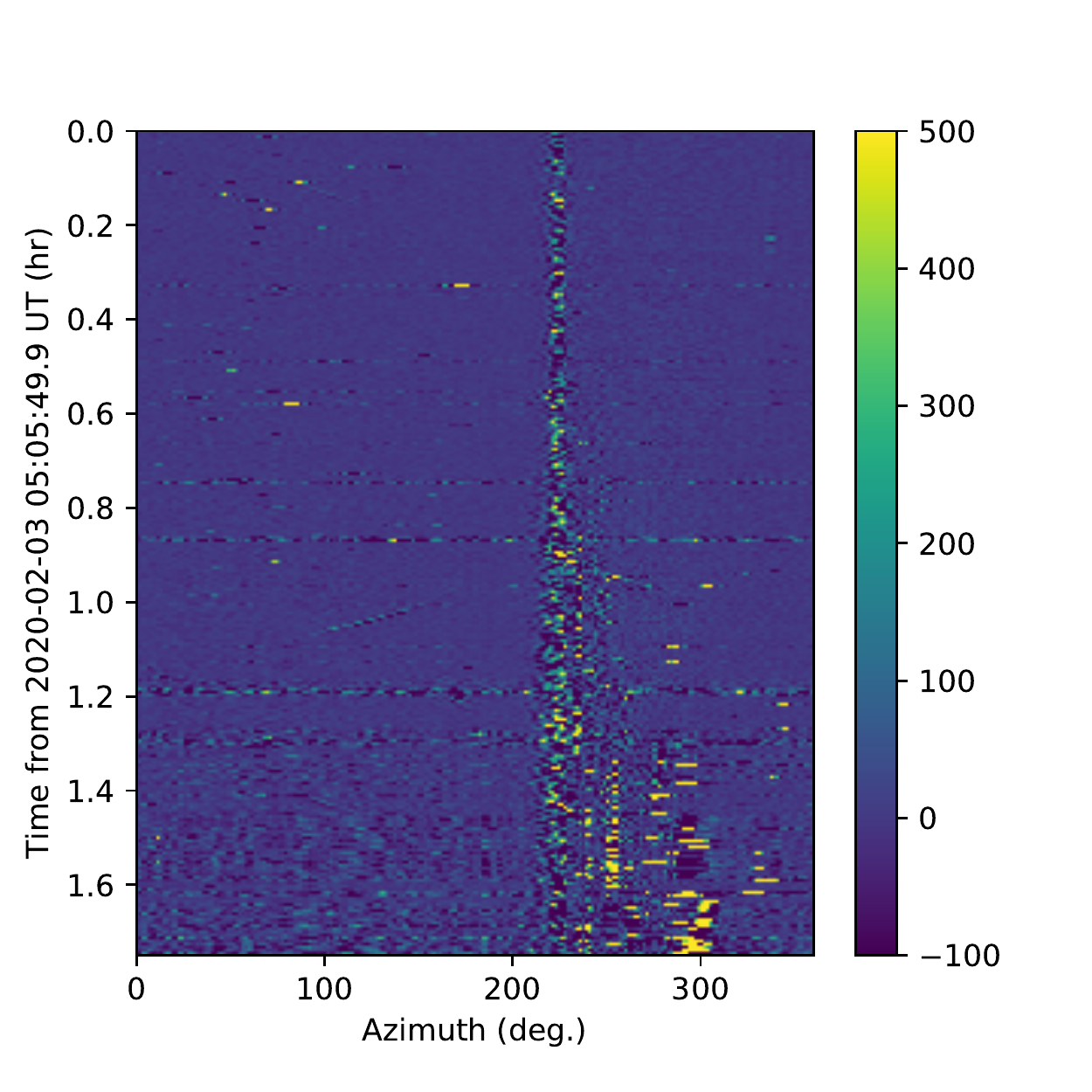}
\caption{The horizon intensity (colourbar units of Jy/beam) extracted from the difference images for a 1.8 hour period on the third day of observations (24 hours after the data shown in Figure \ref{hoxp}), in XX (top) and YY (bottom) polarisations, as a function of azimuth angle (horizontal axis) and time (vertical axis).}
\label{hoxf}
\end{center}
\end{figure}

However, Figure \ref{hoxf} shows the same measurements as in Figure \ref{hoxp} but 24 hours later, in the final 1.8 hours of the observations, showing strikingly different behaviour.  In these data, strong signals are seen apparently propagating around almost the entire western horizon, over periods of 100s to 1000s of seconds.  These signals are attributed to lightning to the west of the MRO and are verified with reference to contemporaneous data from the BIGHORNS antenna, which show broadband signals typical of lightning at the same time (Sokokowski, priv. comm.).  As broadband signals, lightning will affect a much wider bandwidth than just the FM band.

\subsection{Meteor reflections}
\label{subsec:met}

After taking account of false detections associated with bright astronomical radio sources, signals due to direct reception of transmissions from distant population centres, and reflections and/or transmissions from aircraft and satellites (at least those obviously identifiable), we are left with signals that are confined to single time steps and are distributed across the sky.

Overwhelmingly, these signals are likely to be due to FM reflections off meteor trails.  We cannot rule out that some of these signals are due to short duration 'glints' off satellites in Earth orbit.  However, given the results described in \S \ref{subsec:orb}, this is likely to form a very minor contribution.

\begin{figure}[h!]
\begin{center}
\includegraphics[width=0.4\textwidth]{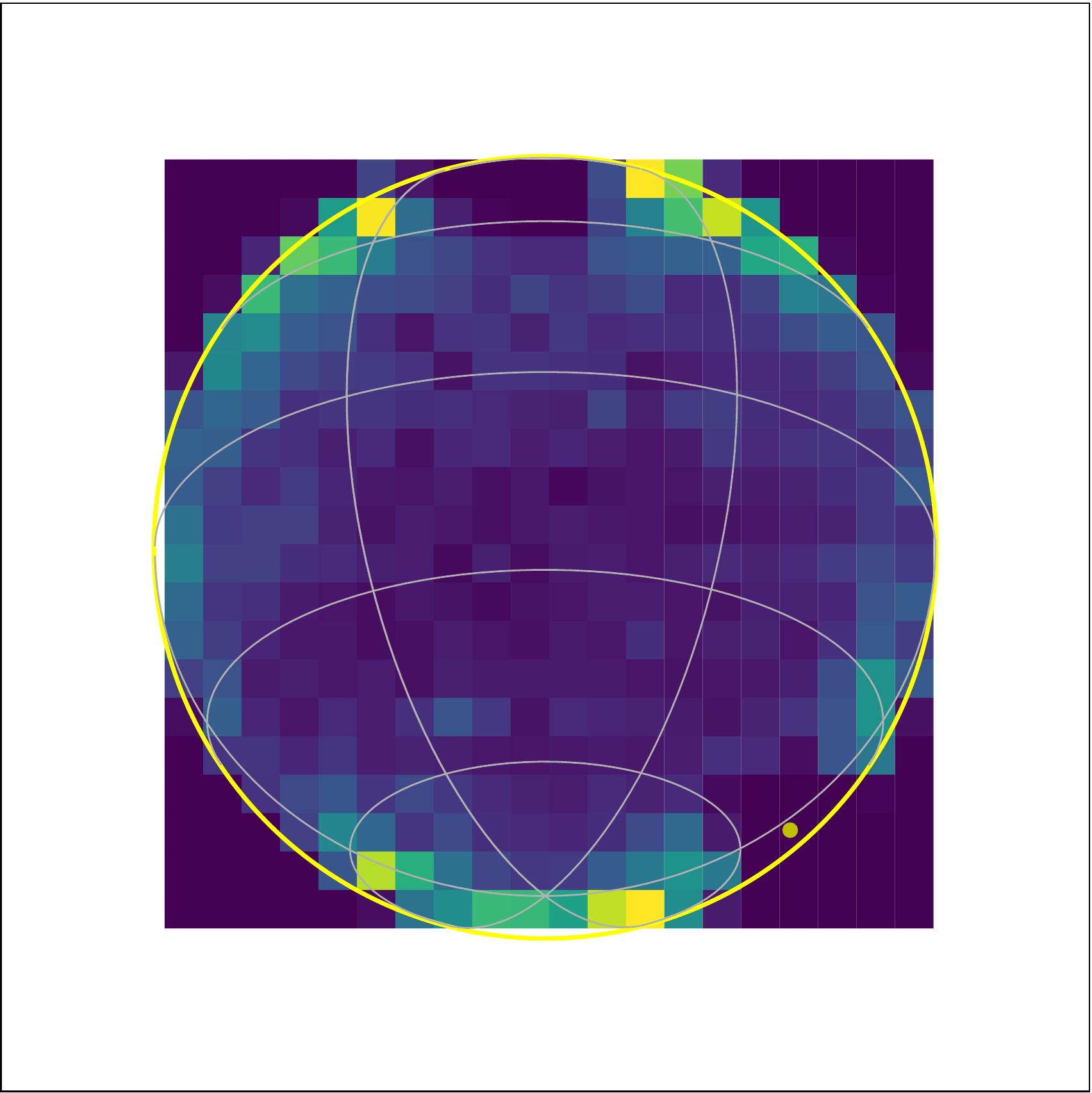}
\caption{The two dimension histogram of the 3,352 detections likely to be reflections from meteor trails, for the XX polarisation.}
\label{met2d}
\end{center}
\end{figure}

\begin{figure}[h!]
\begin{center}
\includegraphics[width=0.45\textwidth]{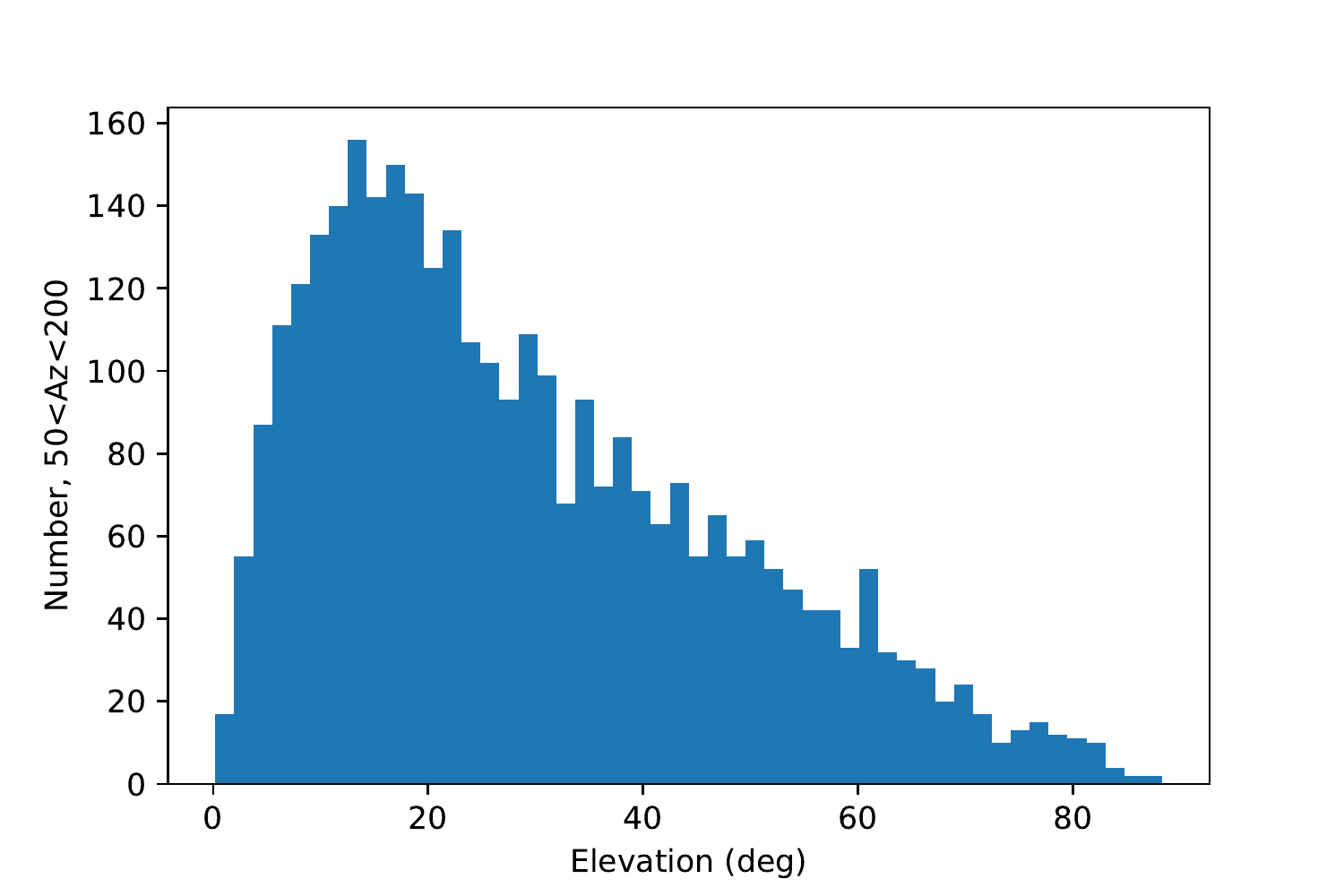}
\caption{The elevation angle dependence of meteor reflections, in the azimuth range $50^{\circ}<Az<200^{\circ}$, for the XX polarisation.}
\label{metaa}
\end{center}
\end{figure}

Figure \ref{met2d} shows the two-dimensional histogram of these events over the sky, for the XX polarisation, with Figure \ref{metaa} showing the distribution of events as a function of elevation (between azimuths of 50${^\circ}$ and 200$^{\circ}$, in order to avoid the significant transmitters on the horizon shown in Figures \ref{hoxp} and \ref{hoxf}).  These distributions are typical of expectations for all-sky meteor radars, as simulated in \citet{Holdsworth2004}.  Within this azimuth range, we detect 3,352 meteors in the XX polaristion and 3,252 in the YY polarisation.  Adjusting these numbers for the full sky, we estimate approximately 8,000 meteors over the three days, or approximately 2,700 meteors per day.

The detailed count rates of meteor reflections as functions of time and position on the sky (not during meteor showers) are functions of the distribution of the six sporadic meteor radiants, the illumination pattern of the radar transmitter(s), and the beam pattern of the receiving antenna system \citep{Cervera2004}.  In the case here, where the illumination is provided by an ensemble of commercial FM transmitters, the illumination pattern is not well known.  However, the estimated daily rates found here are comparable to those found by \citet{Holdsworth2004}.

\begin{figure}[h!]
\begin{center}
\includegraphics[width=0.45\textwidth]{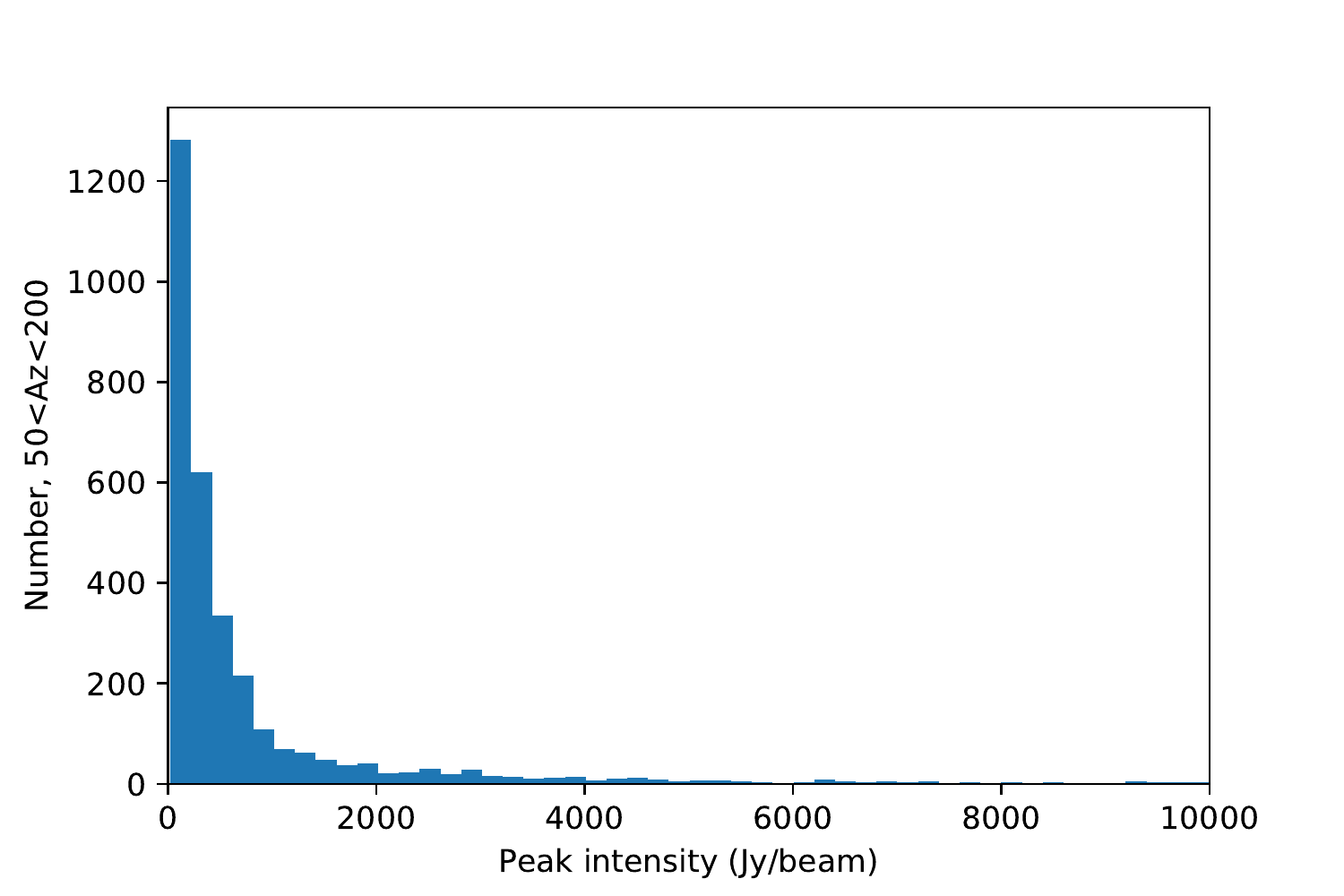}
\caption{Peak intensity of signals due to reflections from meteor trails in the XX polarisation.  The distribution is truncated at 10,000 Jy/beam, due to the very long tail of the distribution that extends to $\sim$1 million Jy/beam.}
\label{metfd}
\end{center}
\end{figure}

Finally, Figure \ref{metfd} shows the distribution of peak intensity for the meteor reflections for the XX polarisation (the YY polarisation results are similar and are not shown).

\subsection{Corrections for antenna beam pattern and bandwidth/time dilution}
\label{subsec:corr}
The data presented above describe the signals entering the EDA2 system via the MWA antennas and thus represent the power injected into the system after modification due to the beam pattern due to the antenna (variable sensitivity across the sky).  In order to consider the signals we have studied here in general terms, for example to consider the effect of these signals on a different antenna type with a different beam pattern, we need to correct the observed apparent intensities with a model for the MWA antenna beam.

\begin{figure}[h!]
\begin{center}
\includegraphics[width=0.22\textwidth]{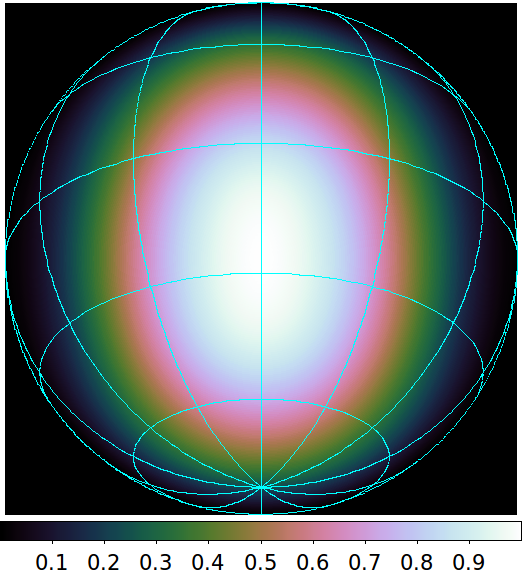}
\includegraphics[width=0.22\textwidth]{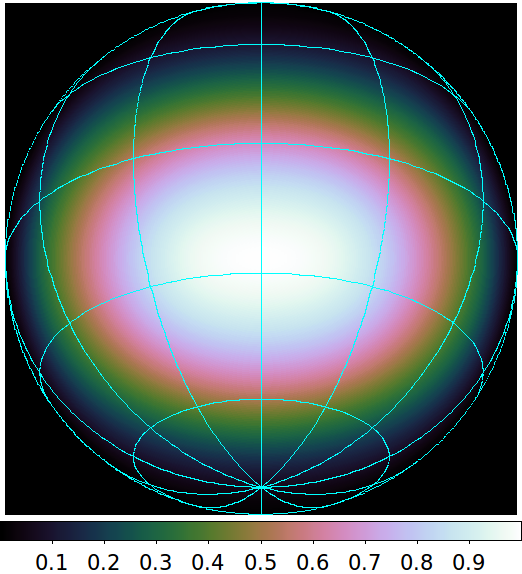}
\caption{X (left) and Y (right) polarisation simulated beam power patterns for an MWA dipole antenna over an infinite ground plane, used to correct observed apparent intensities.  The peak values are normalised to unity at the zenith.}
\label{beams}
\end{center}
\end{figure}

We have used a simulation of the beam pattern for a single MWA dual-polarisation dipole antenna over an infinite ground plane, as used previously to compare the sensitivity expectations to measurements for the first generation EDA instrument by \citet{2017PASA...34...34W}, to correct the observed apparent intensities into calibrated intensities.  The beam patterns for the X and Y polarisations, at the frequency of observation (98 MHz), are shown in Figure \ref{beams}.

Near the horizon, the beam correction is largest and generally most uncertain in simulated beam patterns.  Additionally, mutual coupling effects between antennas in the array will generally produce deviations from the beam expected from a single isolated antenna, resulting in the so-called embedded element pattern.  For the MWA bow-tie antennas used in the EDA2 configuration, these mutual coupling effects are extremely minimal, even toward the horizon and do not represent a significant complication to the calibration.  A comprehensive set of detailed embedded element electromagnetic simulations have been performed to evaluate the EDA2, as part of SKA design activities, to be published by Davidson et al. (2020, in preparation).

For the aircraft reflections, satellite reflections, and meteor reflections described above, we have produced a list of all detections for both polarisations.  These lists contain: date; time (UTC); azimuth; elevation; right ascension; declination; observed peak intensity; and beam-corrected peak intensity.  For horizon transmitters, we construct a list of detections for the Geraldton transmitters only, with the same fields, excluding azimuth and elevation and only for apparent intensities (not beam corrected intensities). The (azimuth, elevation) is fixed at ($-$139,0). These lists are available online, accompanying this paper, and form the survey results that can be used to estimate the impact of RFI at FM frequencies for radio telescopes at the MRO.

Whereas the data for the meteor reflections in Figure \ref{metfd} are not well described by a powerlaw distribution, the beam-corrected data are well described by a powerlaw distribution ($N\propto F^{-\alpha}$) with $\alpha=1.96\pm0.02$, with an almost identical result (within the uncertainties) for the YY polarisation.

The signals recorded by the EDA2 are for an approximate 1 MHz of bandwidth, containing two FM transmitter frequencies, at 98.1 and 98.9 MHz.  It is important to point out that the FM transmitters each have an approximate transmission bandwidth of 180 kHz and thus do not fill the 1 MHz recorded bandwidth.  The peak intensity measurements are therefore diluted by the 1 MHz band.  We cannot accurately disentangle the relative contributions of the individual FM transmissions to the 1 MHz band, but can make an approximate average correction.  

If we assume that both transmitters make an equal contribution to the recorded signals, half the received power can be attributed to each transmitter and the sum of their transmitter bandwidths makes up 40\% of the recorded bandwidth, thereby causing a factor of 2.5 bandwidth dilution.  Thus, an approximation of the peak intensity for each FM transmitter is 2.5/2=1.25 higher than the measured (and beam pattern corrected) peak intensities recorded in the online data accompanying this paper.  Additionally, the amplitude calibration of the array utilised the Sun, which was at 9$^{\circ}$ from the zenith, leading to a small underestimate of the calibrated intensity, approximately 5\%.  Finally, the beam corrections are more uncertain at the horizion and very low elevations than at significant elevations ($\gtrsim10^{\circ}$).  Each of these uncertainties are small in the context of the intensity distributions shown earlier and the propagation effects due to the atmosphere, in the case of low elevation signals.

A temporal dilution of the signals is also likely, but cannot be estimated.

\section{DISCUSSION AND CONCLUSIONS}
\label{sec:dis}
For the first time, this study has spatially and temporally resolved sources of RFI at the MRO, at one frequency within the FM band.  This has allowed a decomposition of RFI sources into well-isolated categories that generally represent the main forms of RFI experienced by well designed radio telescopes (i.e. excluding sources of self-generated RFI by electronic equipment and poor shielding).

At low radio frequencies, even at a remote and radio quiet location, such as the MRO, complex propagation effects allow the reception of ducted and scattered RFI origining at large distances.  In the data presented here, we see all of these effects, making the data a valuable probe of the RFI environment at the MRO in the FM band, as well as the general characteristics of these effects.

The primary sources of scattered RFI we observe are due to reflections off aircraft and meteors, originating from transmitters at large distances.  In both cases, the received power at the telescope reaches into the tens of thousands of Jy for a very significant proportion of the events recorded.  We found that reflections from aircraft are common at the MRO, due to the routine flight paths that cross the MRO on north-south tracks, occupying 13\% of the observing time over a three day period.  
These results are for a limited frequency range, approximately 1 MHz centred at 98.4375 MHz, designed to contain signals from two Geraldton-based FM radio stations.  Therefore, our data represent a worst case in some form, as we have chosen the nearest and strongest transmitters.  However, we would expect qualitatively similar results at other frequencies in the FM band, from more distant transmitters.  Moreover, we would expect similar results at any other frequency in the SKA band where transmitters exist at similar distances and transmission powers.  Thus, the results here are a reasonably general representation of the effects that will be seen at other applicable frequencies.

\subsection{Impact on Epoch of Reionisation experiments}
\label{subsec:eor}

Measurements of this type are important in order to understand the detailed environment in which an interferometer operates, to obtain an overall understanding of the performance of a system for key science goals.  In particular, a significant part of the science mission for the MWA and the low frequency component of the SKA, is the search for signals from the so-called Epoch of Reionisation (EoR), requiring very high sensitivity, high stability of the system in frequency and time, and an exquisite understanding the ``foreground'' signals (all power that enters the signal path that has been produced in the last 13 billion years!)

\begin{figure}[h!]
\begin{center}
\includegraphics[width=0.37\textwidth,angle=270]{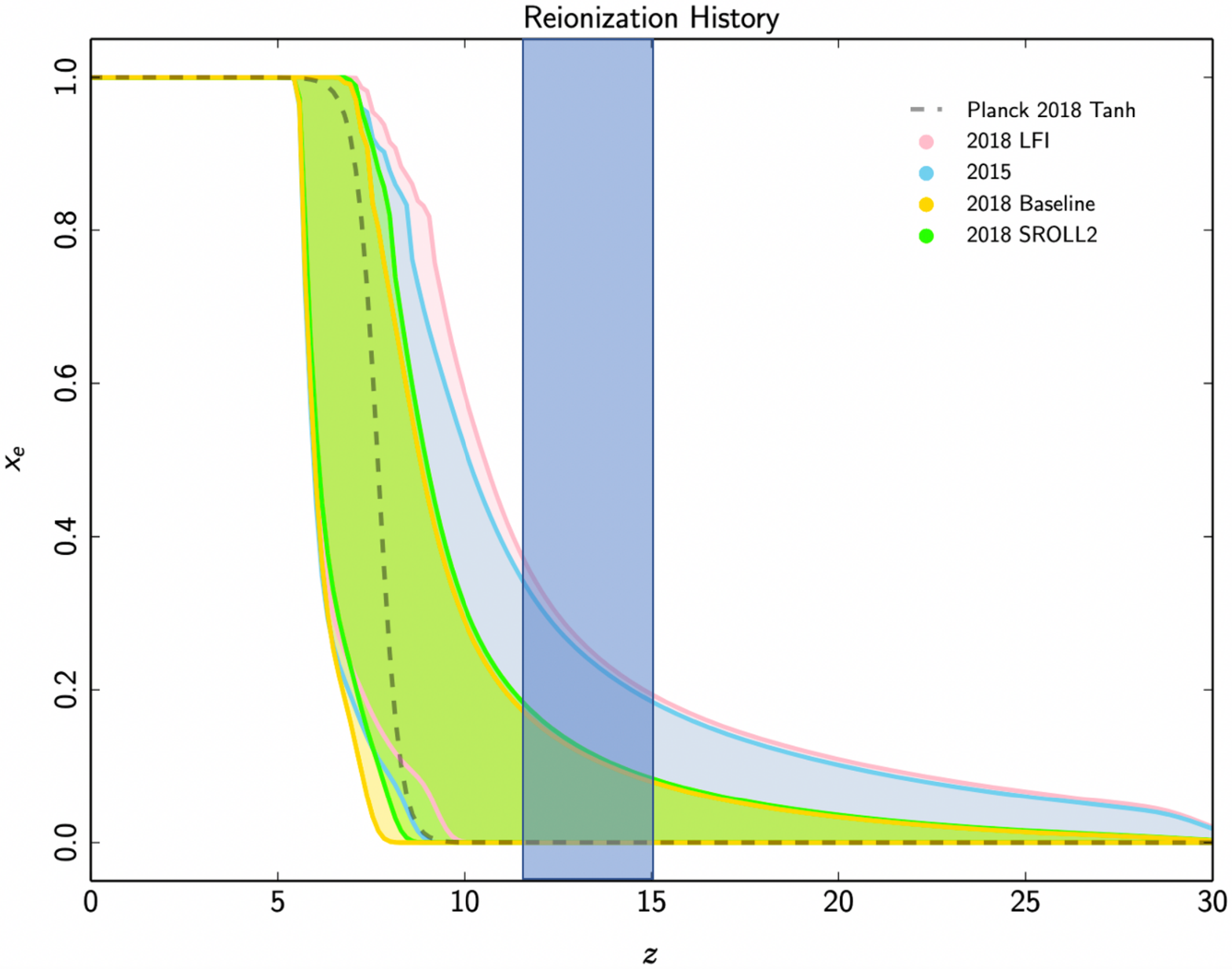}
\caption{Cosmic reionisation history scenarios (ionisation fraction as function of redshift) adapted from \citet{2020arXiv200512222P}.  The green shaded region represents the most recent expectation from Planck data.  The vertical blue shaded block represents the FM band in redshift space.}
\label{eor-planck}
\end{center}
\end{figure}

The most recent estimates of the cosmic reionisation history, based on Planck cosmic microwave background anisotropy data, indicate that later onset and shorter duration reionisation scenarios are favoured \citep{2020arXiv200512222P}.  Figure \ref{eor-planck} shows the currently favoured scenario (2018 SROLL2) for ionisation fraction as a function of redshift, relative to the redshift range corresponding to the FM band relevant to the RFI analysis presented here ($z=\sim12$ -- $\sim15$).  Expectations from Planck continue to place the majority of the evolution in the EoR at lower redshifts, that is higher frequencies than the top of the FM band.  The frequency range of 100 - 200 MHz continues to be strongly favoured to contain the bulk of the EoR signal.

The EoR foreground signals include RFI, which are many orders of magnitude stronger than the astronomical signals, very highly variable in time, and are spatially distributed (as demonstrated in this paper).  How does RFI degrade a system's ability to detect the EoR signals?  \citet{2020arXiv200407819W} finds that narrowband signals of order mJy can challenge the levels of sensitivity required to detect the EoR.  At first glance, given that the RFI signals detected in this paper are in the tens of thousands of Jy, these thresholds do not appear promising.  However, the mJy limits are only relevant to signals after temporal and spatial dilution, and also after efforts have been made to identify and excise RFI from data.

Thus, assessing if a particular observational scenario, such as for the EoR experiment, satisfies the limits derived by \citet{2020arXiv200407819W} is a complex task that requires a detailed model for the response of the telescope, a model that effectively describes where the telescope is pointing and when, and a detailed temporal and spatial model for the RFI environment.  This analysis is beyond the scope of this paper, but our work does provide one of these crucial ingredients, that has been missing until now, an empirical measurement of the temporally and spatially resolved RFI environment.  

As a result of our analysis, we have generated separate lists for meteor reflections, satellite reflections, aircraft reflections, and horizon transmitters in both polarisations, and recording time of event, azimuth and elevation (and corresponding Right Ascension and Declination), and peak intensity, over an approximate 72 hour period (both observed and corrected for antenna beam pattern effects).  We will utilise this information, using the methods in \citet{2020arXiv200407819W} to evaluate the detailed impact of RFI on the EoR experiment for instruments based at the MRO, and will report the results in a future publication.

These lists are available as online data accompanying this paper.

\subsection{Impacts on ground-based astronomy by satellites}
\label{subsec:satimpact}

With the increasing utilisation of the space environment for a variety of purposes, supported by new technologies that allow small, inexpensive satellites and mass launch capabilities, concerns regarding the impact on ground-based astronomy are also increasing \citep{mcdowell2020low,gallozzi2020concerns,hainaut2020impact,mallama2020flat}.  A lot of the focus to date has been on the impact to ground-based optical telescopes, the most prominent example being with respect to the StarLink constellation of satellites.

However, all small satellites have uplink/downlink communications systems that utilise radio frequencies.  Popular frequencies include those within the 2 m amateur band, with downlink transmitters in the 144 - 148 MHz range.  This band is in the middle of the frequency range of the MWA and the low frequency SKA.  Therefore the performance of these transmitters is of very high interest in the radio astronomy community.  With inexpensive electronics, potentially limited testing, and the potential for malfunction on orbit, these transmitters can pose a challenge for telescopes such as the SKA.  A recent study shows that communications subsystems contribute to 14\%, 16\%, and 29\% of cubesat mission failures within the first 0 (Dead On Arrival: DOA), 30, and 90 days on orbit \citep{inproceedings}.  Thus, these systems are prone to malfunctions of various types.

Our results in \S \ref{subsec:orb} indicate that the small satellites MAX VALIER SAT and FLOCK 3P 71 may be transmitting out of their designated transmission bands and that the BGUSAT satellite is very likely to be transmitting out of band.  These results add to previous out of band detections of DUCHIFAT-1 and UKube-1 by \citet{2020arXiv200201674P} and \citet{2018MNRAS.477.5167Z}.  Thus, there is a growing list of small satellites for which there is evidence for significant out of band transmission.

It is not straightforward to determine the characterisation of out of band or spurious transmissions from a satellite, or determine conclusively if a given satellite is violating its communications license, according to documentation of the International Telecommunications Union (ITU)\footnote{https://www.itu.int/dms\_pubrec/itu-r/rec/sm/R-REC-SM.1541-6-201508-I\!\!PDF-E.pdf}.  However, even if out of band transmissions are within ITU requirements, Table \ref{satdet} shows that signals in the thousands of Jy are seen, which are very significant in astronomical terms.

Moreover, the available documentation for many cubesats is patchy and information is difficult to source.  Several organisations and individuals attempt to maintain compilations of available information regarding the characteristics of satellite transmitters, such as at https://db.satnogs.org/ (an open source database for satellite transmitters) or at https://www.klofas.com/comm-table/table.pdf.  However, these resources are not complete or official in nature.

According to https://www.klofas.com/comm-table/table.pdf, both DUCHIFAT-1 and UKube-1 utilise the ISIS TRXUV transceiver, operating with downlink frequencies near 145 MHz\footnote{https://www.isispace.nl/wp-content/uploads/2016/02/ \\ ISIS\_TRXUV\_Transceiver\_Brochure\_v.10.1.pdf}.  The MAX VALIER SAT may also utilise the same transceiver\footnote{https://www.pe0sat.vgnet.nl/2012/max-valier/}.  The FLOCK 3P series of satellites utilises a 401.3 MHz downlink frequency on hardware developed by the constellation owner (an open source version of this hardware is also popular for cubesat development: https://www.planet.com/pulse/planet-openlst-radio-solution-for-cubesats/).  For BGUSAT, no information on the transmitter hardware is available (although the frequencies are known - 145 MHz).

While the level of information available and the number of our observations is small, it is premature to cast aspersions on any particular electronics or cubesat configurations.  Further observations and measurements of satellites on orbit are required.  For example, according to https://www.klofas.com/comm-table/table.pdf a further 10 satellites are listed as having the same transeiver as aboard DUCHIFAT-1 and UKube-1 (NORAD IDs: 38083; 39438; 39427; 39428; 40025; 40032; 40024; 40057; 42731; and 43131).  Observations of sets of satellites with similar transceiver configurations may be useful to determine any trends in performance.

\begin{acknowledgements}
We thank the anonymous referee for constructive comments that improved aspects of the paper.  This research has made use of the NASA/IPAC Extragalactic Database (NED), which is funded by the National Aeronautics and Space Administration and operated by the California Institute of Technology.
This scientific work makes use of the Murchison Radio-astronomy Observatory, operated by CSIRO. We acknowledge the Wajarri Yamatji people as the traditional owners of the Observatory site.  This research made use of Photutils, an Astropy package for
detection and photometry of astronomical sources \citep{Bradley_2019_2533376}.  The acquisition system was designed and purchased by INAF/Oxford University and the RX chain was design by INAF, as part of the SKA design and prototyping program.
\end{acknowledgements}

\bibliographystyle{pasa-mnras}
\bibliography{custom}

\begin{thebibliography}{}
\makeatletter
\relax
\def\mn@urlcharsother{\let\do\@makeother \do\$\do\&\do\#\do\^\do\_\do\%\do\~}
\definecolor{darkblue}{rgb}{0,0,0.597656}
\def\mndoi{\begingroup\mn@urlcharsother \@ifnextchar [ {\mndoi@} {\mndoi@[]}}
\def\mndoi@[#1]#2{\def\@tempa{#1}\ifx\@tempa\@empty \href
  {http://dx.doi.org/#2} {\textcolor{darkblue}{doi:#2}}\else \href
  {http://dx.doi.org/#2} {\textcolor{darkblue}{#1}}\fi \endgroup}
\def\mn@eprint#1#2{\mn@eprint@#1:#2::\@nil}
\def\mn@eprint@arXiv#1{\href {http://arxiv.org/abs/#1} {{\tt arXiv:#1}}}
\def\mn@eprint@dblp#1{\href {http://dblp.uni-trier.de/rec/bibtex/#1.xml}
  {dblp:#1}}
\def\mn@eprint@#1:#2:#3:#4\@nil{\def\@tempa {#1}\def\@tempb {#2}\def\@tempc
  {#3}\ifx \@tempc \@empty \let \@tempc \@tempb \let \@tempb \@tempa \fi \ifx
  \@tempb \@empty \def\@tempb {arXiv}\fi \@ifundefined
  {mn@eprint@\@tempb}{\@tempb:\@tempc}{\expandafter \expandafter \csname
  mn@eprint@\@tempb\endcsname \expandafter{\@tempc}}}

\bibitem[\protect\citeauthoryear{Benz}{Benz}{2009}]{sun_model}
Benz A.~O.,  2009, 4.1.1.6 Radio emission of the quiet Sun: Datasheet from
  Landolt-B{\"o}rnstein - Group VI Astronomy and Astrophysics
  {\textperiodcentered} Volume 4B: ``Solar System'' in SpringerMaterials
  (https://doi.org/10.1007/978-3-540-88055-4{\_}5),
  \mndoi{10.1007/978-3-540-88055-4_5}, \url
  {https://materials.springer.com/lb/docs/sm_lbs_978-3-540-88055-4_5}

\bibitem[\protect\citeauthoryear{Bradley et~al.,}{Bradley
  et~al.}{2019}]{Bradley_2019_2533376}
Bradley L.,  et~al., 2019, astropy/photutils: v0.6,
  \mndoi{10.5281/zenodo.2533376}, \url {https://doi.org/10.5281/zenodo.2533376}

\bibitem[\protect\citeauthoryear{Cervera}{Cervera}{2004}]{Cervera2004}
Cervera M.~A.,  2004, \mndoi [Journal of Geophysical Research]
  {10.1029/2004ja010450}, 109

\bibitem[\protect\citeauthoryear{{Clark}, {La Plante}  \& {Greenhill}}{{Clark}
  et~al.}{2011}]{xGPU}
{Clark} M.~A.,  {La Plante} P.~C.,   {Greenhill} L.~J.,  2011, arXiv e-prints,
  \href {https://ui.adsabs.harvard.edu/abs/2011arXiv1107.4264C} {p.
  arXiv:1107.4264}

\bibitem[\protect\citeauthoryear{{Comoretto} et~al.,}{{Comoretto}
  et~al.}{2017}]{2017JAI.....641015C}
{Comoretto} G.,  et~al., 2017, \mndoi [Journal of Astronomical Instrumentation]
  {10.1142/S2251171716410154}, \href
  {https://ui.adsabs.harvard.edu/abs/2017JAI.....641015C} {6, 1641015}

\bibitem[\protect\citeauthoryear{{Furlanetto}, {Oh}  \& {Briggs}}{{Furlanetto}
  et~al.}{2006}]{2006PhR...433..181F}
{Furlanetto} S.~R.,  {Oh} S.~P.,   {Briggs} F.~H.,  2006, \mndoi [\physrep]
  {10.1016/j.physrep.2006.08.002}, \href
  {https://ui.adsabs.harvard.edu/abs/2006PhR...433..181F} {433, 181}

\bibitem[\protect\citeauthoryear{Gallozzi, Scardia  \& Maris}{Gallozzi
  et~al.}{2020}]{gallozzi2020concerns}
Gallozzi S.,  Scardia M.,   Maris M.,  2020, Concerns about ground based
  astronomical observations: a step to safeguard the astronomical sky
  (\mn@eprint {arXiv} {2001.10952})

\bibitem[\protect\citeauthoryear{{Greisen}}{{Greisen}}{2019}]{uvfits}
{Greisen} E.~W.,  2019, Aips memo series, AIPS FITS File Format.
AIPS Memo 117

\bibitem[\protect\citeauthoryear{Hainaut \& Williams}{Hainaut \&
  Williams}{2020}]{hainaut2020impact}
Hainaut O.~R.,  Williams A.~P.,  2020, Astronomy \& Astrophysics, 636, A121

\bibitem[\protect\citeauthoryear{{Hennessy} et~al.,}{{Hennessy}
  et~al.}{2019}]{8835821}
{Hennessy} B.,  et~al., 2019, in 2019 IEEE Radar Conference (RadarConf).
  pp~1--6, \mndoi{10.1109/RADAR.2019.8835821}

\bibitem[\protect\citeauthoryear{{Hennessy}, {Rutten}, {Tingay}  \&
  {Young}}{{Hennessy} et~al.}{2020}]{2020arXiv200303947H}
{Hennessy} B.,  {Rutten} M.,  {Tingay} S.,   {Young} R.,  2020, arXiv e-prints,
  \href {https://ui.adsabs.harvard.edu/abs/2020arXiv200303947H} {p.
  arXiv:2003.03947}

\bibitem[\protect\citeauthoryear{Holdsworth, Tsutsumi, Reid, Nakamura  \&
  Tsuda}{Holdsworth et~al.}{2004}]{Holdsworth2004}
Holdsworth D.~A.,  Tsutsumi M.,  Reid I.~M.,  Nakamura T.,   Tsuda T.,  2004,
  \mndoi [Radio Science] {10.1029/2003rs003026}, 39, n/a

\bibitem[\protect\citeauthoryear{{Johnston} et~al.,}{{Johnston}
  et~al.}{2008}]{2008ExA....22..151J}
{Johnston} S.,  et~al., 2008, \mndoi [Experimental Astronomy]
  {10.1007/s10686-008-9124-7}, \href
  {https://ui.adsabs.harvard.edu/abs/2008ExA....22..151J} {22, 151}

\bibitem[\protect\citeauthoryear{Langer \& Bouwmeester}{Langer \&
  Bouwmeester}{2016}]{inproceedings}
Langer M.,  Bouwmeester J.,  2016.

\bibitem[\protect\citeauthoryear{Mallama}{Mallama}{2020}]{mallama2020flat}
Mallama A.,  2020, arXiv preprint arXiv:2003.07805

\bibitem[\protect\citeauthoryear{McDowell}{McDowell}{2020}]{mcdowell2020low}
McDowell J.~C.,  2020, The Low Earth Orbit Satellite Population and Impacts of
  the SpaceX Starlink Constellation (\mn@eprint {arXiv} {2003.07446})

\bibitem[\protect\citeauthoryear{Monroe}{Monroe}{2018}]{https://doi.org/10.7907/25dp-j474}
Monroe R.~M.,  2018, Gigahertz Bandwidth and Nanosecond Timescales: New
  Frontiers in Radio Astronomy Through Peak Performance Signal Processing,
  \mndoi{10.7907/25DP-J474}, \url
  {https://resolver.caltech.edu/CaltechTHESIS:06042018-004220017}

\bibitem[\protect\citeauthoryear{{Naldi} et~al.,}{{Naldi}
  et~al.}{2017}]{2017JAI.....641014N}
{Naldi} G.,  et~al., 2017, \mndoi [Journal of Astronomical Instrumentation]
  {10.1142/S2251171716410142}, \href
  {https://ui.adsabs.harvard.edu/abs/2017JAI.....641014N} {6, 1641014}

\bibitem[\protect\citeauthoryear{{Offringa} et~al.,}{{Offringa}
  et~al.}{2013}]{2013A&A...549A..11O}
{Offringa} A.~R.,  et~al., 2013, \mndoi [\aap] {10.1051/0004-6361/201220293},
  \href {https://ui.adsabs.harvard.edu/abs/2013A&A...549A..11O} {549, A11}

\bibitem[\protect\citeauthoryear{{Offringa} et~al.,}{{Offringa}
  et~al.}{2015}]{2015PASA...32....8O}
{Offringa} A.~R.,  et~al., 2015, \mndoi [\pasa] {10.1017/pasa.2015.7}, \href
  {https://ui.adsabs.harvard.edu/abs/2015PASA...32....8O} {32, e008}

\bibitem[\protect\citeauthoryear{{Palmer} et~al.,}{{Palmer}
  et~al.}{2017}]{7944483}
{Palmer} J.~E.,  et~al., 2017, in 2017 IEEE Radar Conference (RadarConf). pp
  1715--1720, \mndoi{10.1109/RADAR.2017.7944483}

\bibitem[\protect\citeauthoryear{{Paoletti}, {Hazra}, {Finelli}  \&
  {Smoot}}{{Paoletti} et~al.}{2020}]{2020arXiv200512222P}
{Paoletti} D.,  {Hazra} D.~K.,  {Finelli} F.,   {Smoot} G.~F.,  2020, arXiv
  e-prints, \href {https://ui.adsabs.harvard.edu/abs/2020arXiv200512222P} {p.
  arXiv:2005.12222}

\bibitem[\protect\citeauthoryear{{Prabu}, {Hancock}, {Zhang}  \&
  {Tingay}}{{Prabu} et~al.}{2020a}]{2020arXiv200201674P}
{Prabu} S.,  {Hancock} P.~J.,  {Zhang} X.,   {Tingay} S.~J.,  2020a, arXiv
  e-prints, \href {https://ui.adsabs.harvard.edu/abs/2020arXiv200201674P} {p.
  arXiv:2002.01674}

\bibitem[\protect\citeauthoryear{{Prabu}, {Hancock}, {Zhang}  \&
  {Tingay}}{{Prabu} et~al.}{2020b}]{2020PASA...37...10P}
{Prabu} S.,  {Hancock} P.~J.,  {Zhang} X.,   {Tingay} S.~J.,  2020b, \mndoi
  [\pasa] {10.1017/pasa.2020.1}, \href
  {https://ui.adsabs.harvard.edu/abs/2020PASA...37...10P} {37, e010}

\bibitem[\protect\citeauthoryear{{Sault}, {Teuben}  \& {Wright}}{{Sault}
  et~al.}{1995}]{1995ASPC...77..433S}
{Sault} R.~J.,  {Teuben} P.~J.,   {Wright} M.~C.~H.,  1995, in {Shaw} R.~A.,
  {Payne} H.~E.,   {Hayes} J.~J.~E.,  eds,  Astronomical Society of the Pacific
  Conference Series Vol. 77, Astronomical Data Analysis Software and Systems
  IV. p.~433 (\mn@eprint {arXiv} {astro-ph/0612759})

\bibitem[\protect\citeauthoryear{{Sokolowski} et~al.,}{{Sokolowski}
  et~al.}{2015}]{2015PASA...32....4S}
{Sokolowski} M.,  et~al., 2015, \mndoi [\pasa] {10.1017/pasa.2015.3}, \href
  {https://ui.adsabs.harvard.edu/abs/2015PASA...32....4S} {32, e004}

\bibitem[\protect\citeauthoryear{{Sokolowski}, {Wayth}  \&
  {Lewis}}{{Sokolowski} et~al.}{2016}]{2016arXiv161004696S}
{Sokolowski} M.,  {Wayth} R.~B.,   {Lewis} M.,  2016, arXiv e-prints, \href
  {https://ui.adsabs.harvard.edu/abs/2016arXiv161004696S} {p. arXiv:1610.04696}

\bibitem[\protect\citeauthoryear{{Sokolowski}, {Wayth}  \&
  {Ellement}}{{Sokolowski} et~al.}{2017}]{2017rfi..confE...1S}
{Sokolowski} M.,  {Wayth} R.~B.,   {Ellement} T.,  2017, in Radio Frequency
  Interference (RFI. p. 7833541, \mndoi{10.1109/RFINT.2016.7833541}

\bibitem[\protect\citeauthoryear{{Stetson}}{{Stetson}}{1987}]{1987PASP...99..191S}
{Stetson} P.~B.,  1987, \mndoi [\pasp] {10.1086/131977}, \href
  {https://ui.adsabs.harvard.edu/abs/1987PASP...99..191S} {99, 191}

\bibitem[\protect\citeauthoryear{{Tingay} et~al.,}{{Tingay}
  et~al.}{2013a}]{2013PASA...30....7T}
{Tingay} S.~J.,  et~al., 2013a, \mndoi [Publications of the Astronomical
  Society of Australia] {10.1017/pasa.2012.007}, \href
  {http://adsabs.harvard.edu/abs/2013PASA...30....7T} {30, e007}

\bibitem[\protect\citeauthoryear{Tingay et~al.,}{Tingay
  et~al.}{2013b}]{Tingay2013OnFeasibility}
Tingay S.~J.,  et~al., 2013b, \mndoi [Astronomical Journal]
  {10.1088/0004-6256/146/4/103}, 146

\bibitem[\protect\citeauthoryear{{Wayth} et~al.,}{{Wayth}
  et~al.}{2017}]{2017PASA...34...34W}
{Wayth} R.,  et~al., 2017, \mndoi [\pasa] {10.1017/pasa.2017.27}, \href
  {https://ui.adsabs.harvard.edu/abs/2017PASA...34...34W} {34, e034}

\bibitem[\protect\citeauthoryear{{Wayth} et~al.,}{{Wayth}
  et~al.}{2018}]{2018PASA...35...33W}
{Wayth} R.~B.,  et~al., 2018, \mndoi [\pasa] {10.1017/pasa.2018.37}, \href
  {https://ui.adsabs.harvard.edu/abs/2018PASA...35...33W} {35, 33}

\bibitem[\protect\citeauthoryear{{Wilensky}, {Barry}, {Morales}, {Hazelton}  \&
  {Byrne}}{{Wilensky} et~al.}{2020}]{2020arXiv200407819W}
{Wilensky} M.~J.,  {Barry} N.,  {Morales} M.~F.,  {Hazelton} B.~J.,   {Byrne}
  R.,  2020, arXiv e-prints, \href
  {https://ui.adsabs.harvard.edu/abs/2020arXiv200407819W} {p. arXiv:2004.07819}

\bibitem[\protect\citeauthoryear{{Zhang} et~al.,}{{Zhang}
  et~al.}{2018}]{2018MNRAS.477.5167Z}
{Zhang} X.,  et~al., 2018, \mndoi [\mnras] {10.1093/mnras/sty930}, \href
  {https://ui.adsabs.harvard.edu/abs/2018MNRAS.477.5167Z} {477, 5167}

\makeatother
\end{thebibliography}

\end{document}